\documentclass[12pt,english]{article}
\pdfoutput=1
\usepackage[T1]{fontenc}
\usepackage[latin9]{inputenc}
\usepackage{geometry}
\usepackage{amsmath}
\usepackage{graphicx}
\usepackage[authoryear]{natbib}
\usepackage{setspace}

\makeatletter

\providecommand{\tabularnewline}{\\}

\usepackage{amsthm}
\newtheorem{thm}{Theorem}

\newtheorem{prop}{Proposition}
\newtheorem{definition}{Definition}
\usepackage[ruled]{algorithm2e}
\usepackage{multirow}
\usepackage{dsfont}
\graphicspath{{./Graphics/}}

\usepackage{babel}
\makeatother

\def\b1{{\bf 1}}

\def\bX{{\bf X}}

\def\bI{{\bf I}}

\def\by{{\bf y}}
\def\bbeta{{\mbox{\boldmath $\beta$}}}

\def\sign{{\rm sign}}

\def\sign{{\rm sign}}

\begin{document}

\title{A path algorithm for the Fused Lasso Signal Approximator}

\author{Holger Hoefling\thanks{Email: hhoeflin@gmail.com} \\ Stanford University}

\maketitle
\begin{abstract}
The Lasso is a very well known penalized regression model, which adds
an $L_{1}$ penalty with parameter $\lambda_{1}$ on the coefficients
to the squared error loss function. The Fused Lasso extends this model
by also putting an $L_{1}$ penalty with parameter $\lambda_{2}$
on the difference of neighboring coefficients, assuming there is a
natural ordering. In this paper, we develop a fast path algorithm for solving
the Fused Lasso Signal Approximator that computes the solutions for all values of
$\lambda_1$ and $\lambda_2$. In the supplement, we also give an algorithm
for the general Fused Lasso for the case with predictor matrix $\bX \in \mathds{R}^{n \times p}$ with 
$\text{rank}(\bX)=p$. 
\end{abstract}

\section{Introduction}

In recent years, many regression procedures have been proposed that
use penalties on the regression coefficients in order to achieve sparseness
or shrink them towards zero. One of the most widely known procedures
of this type is the Lasso (see \citet{Tibshirani1996}), which minimizes
the loss function
\[
\frac{1}{2}(\by-\bX\bbeta)^{T}(\by-\bX\bbeta)+\lambda_{1}\sum_{i=1}^{p}|\beta_{i}|.
\]
Here, $\by \in \mathds{R}^n$ is the response vector, $\bX \in \mathds{R}^{n\times p}$ is the matrix of predictors and $\bbeta \in \mathds{R}^p$ the coefficient vector. 
Several years after the original Lasso paper was published, the LARS
algorithm was developed (see \citet{Efron2002}), which after a small
adjustment gives the whole solution path of the Lasso for the penalty
parameter $\lambda_{1}$ with the computational complexity of an ordinary least squares problem.
Subsequently, path algorithms for several other regression methods
were developed as well, for example for generalized linear models
(see \citet{Park2007}) or the SVM (see \citet{Hastie2004}) among
others. A more general treatment of conditions under which the solution
paths are piecewise linear can be found in \citet{Rosset2007}. 

An example of an extension of the Lasso is the Fused Lasso introduced
in \citet{Tibshirani2005}. For the Fused Lasso, it is assumed that
there is some natural ordering of the coefficients (e.g. each coefficient
corresponds to a position on a straight line). If coefficients in
the true model are closely related to their neighbors, we can exploit
this by placing an additional penalty on the differences of neighboring
coefficients. Several different choices for these penalties on neighboring
coefficients are possible and in the case of the Fused Lasso, an $L_{1}$
penalty is being used. The resulting loss function is then
\[
\frac{1}{2}(\by-\bX\bbeta)^{T}(\by-\bX\bbeta)+\lambda_{1}\sum_{i=1}^{p}|\beta_{i}|+\lambda_{2}\sum_{i=1}^{p-1}|\beta_{i}-\beta_{i+1}|.
\]
The second penalty with parameter $\lambda_{2}$ shrinks neighboring coefficients towards each other. 
Just as the $L_{1}$ penalty on the absolute values $|\beta_{i}|$ for the Lasso encourages sparseness,
the penalty on $|\beta_{i}-\beta_{i+1}|$ tends to set neighboring
penalties exactly equal to each other. As such, the method is especially
suitable for coefficients that are constant for an interval and change
in jumps.

\begin{figure}[t]
\begin{centering}
\includegraphics[width=8cm]{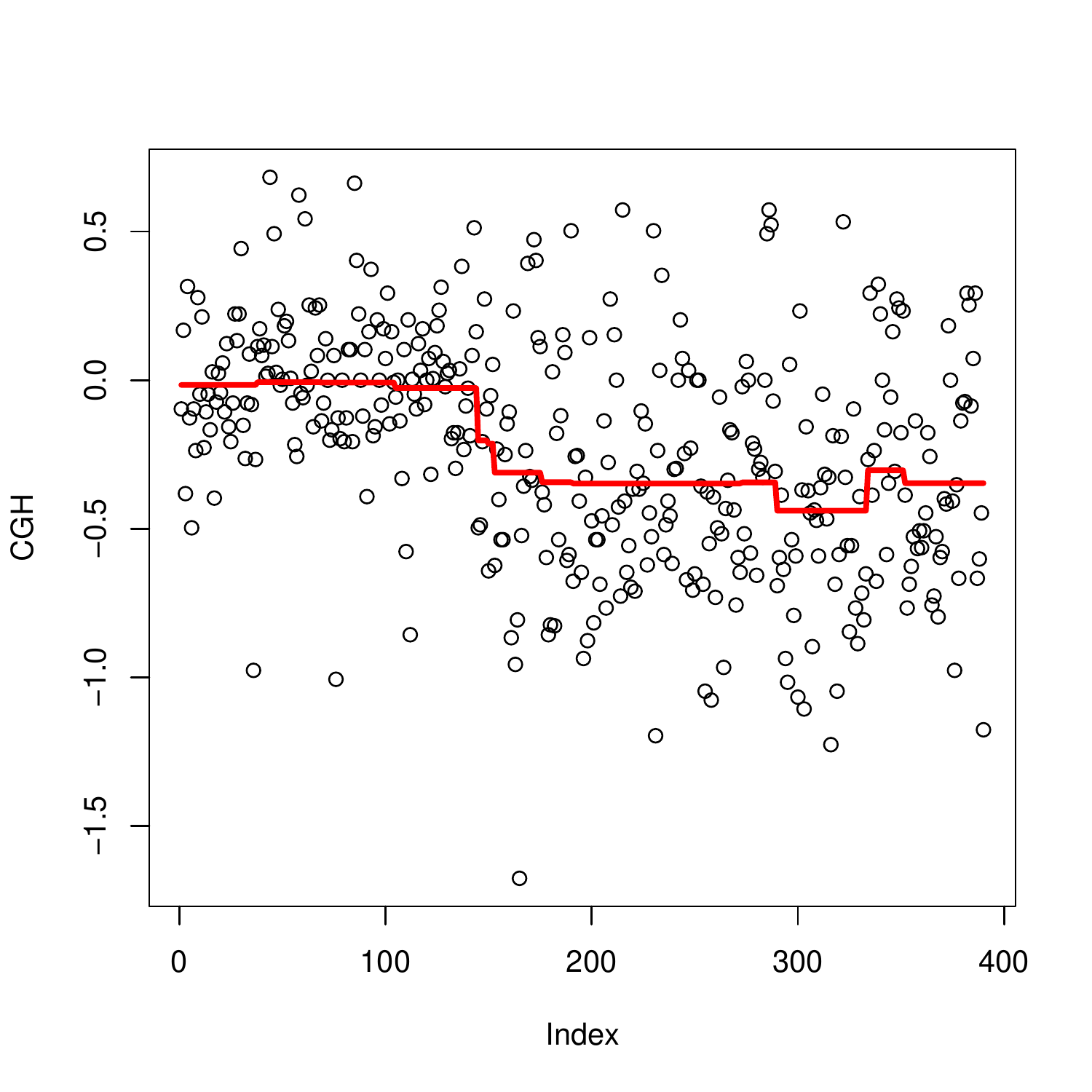}
\caption[Example for 1-dim. FLSA]{Example using the one-dimensional Fused Lasso Signal Approximator
on lung cancer CGH data.}
\label{fig:LungCancerExample}
\end{centering}
\end{figure}

In this article, we first want to concentrate on the most widely used case for this
method, the Fused Lasso Signal Approximator (FLSA). In the FLSA, we assume that we have $\bX=\bI$ as
the predictor matrix. One example for this would be comparative genomic hybridization (CGH)
or chromosomal microarray analysis (CMA) data. CGH is a method that
identifies DNA copy number gains and losses on chromosomes by making
two color fluorescence in situ hybridization at various points of
the chromosomes. In this technique, normal and tumor DNA are labeled with
fluorescent dyes (e.g. red and green) and using a microarray analysis, regions of increased or decreased
fluorescence of one color compared to the other can be identified,
indicating gains or losses of DNA at this place of the chromosome. 
As usual with this type of data, it is very noisy. Therefore, we seek to exploit that
gains or losses typically appear for whole regions in the genome and that these changes usually occur in jumps. We can do this by penalizing differences of neighboring
coefficients and therefore decrease the noise in the data and improve
estimation. In this case, we use the one-dimensional
Fused Lasso Signal Approximator (FLSA), for which the loss function
is 
\[
L(\by,\bbeta)=\frac{1}{2}\sum_{i=1}^{n}(y_{i}-\beta_{i})^{2}+\lambda_{1}\sum_{i=1}^{n}|\beta_{i}|+\lambda_{2}\sum_{i=1}^{n-1}|\beta_{i}-\beta_{i+1}|.
\]
Every coefficient $\beta_{i}$ is an estimate of the measurement $y_{i}$
taken at position i (which we assume to be ordered along the chromosome).
Apart from the Lasso penalty $\lambda_{1}\sum_{i=1}^{n}|\beta_{i}|$, the additional
penalty placed on the difference between neighboring coefficients
is $\lambda_{2}\sum_{i=1}^{n-1}|\beta_{i}-\beta_{i+1}|$. An example
of CGH measurements in lung cancer can be seen in Figure \ref{fig:LungCancerExample}.
The red line are the estimates for penalty parameters $\lambda_{1}=0$
and $\lambda_{2}=2$. We can see that starting around measurement
150, the CGH results are on average below 0, indicating a loss of
DNA in this region. 

\begin{figure}[t]
\begin{centering}
\includegraphics[width=4cm]{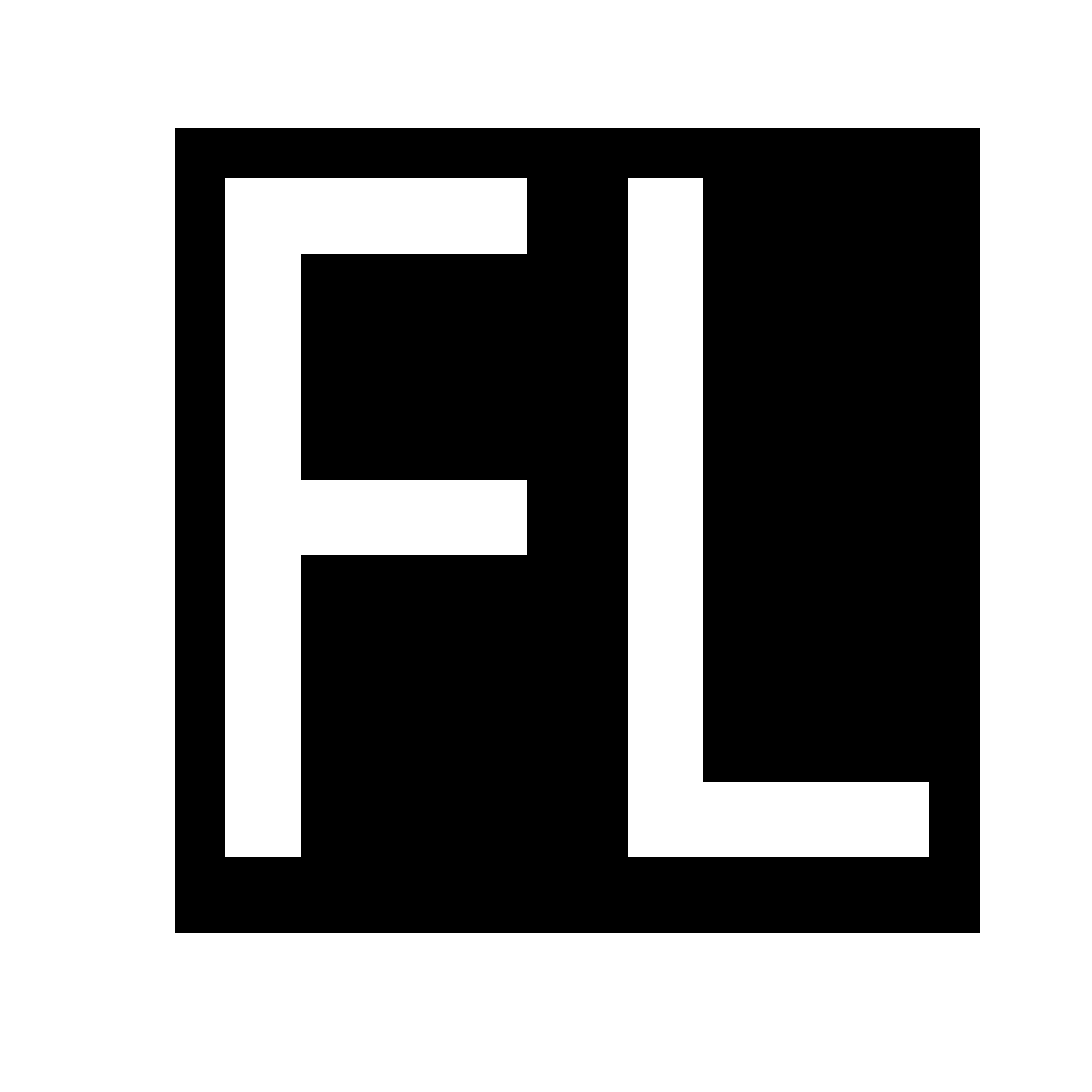}\includegraphics[width=4cm]{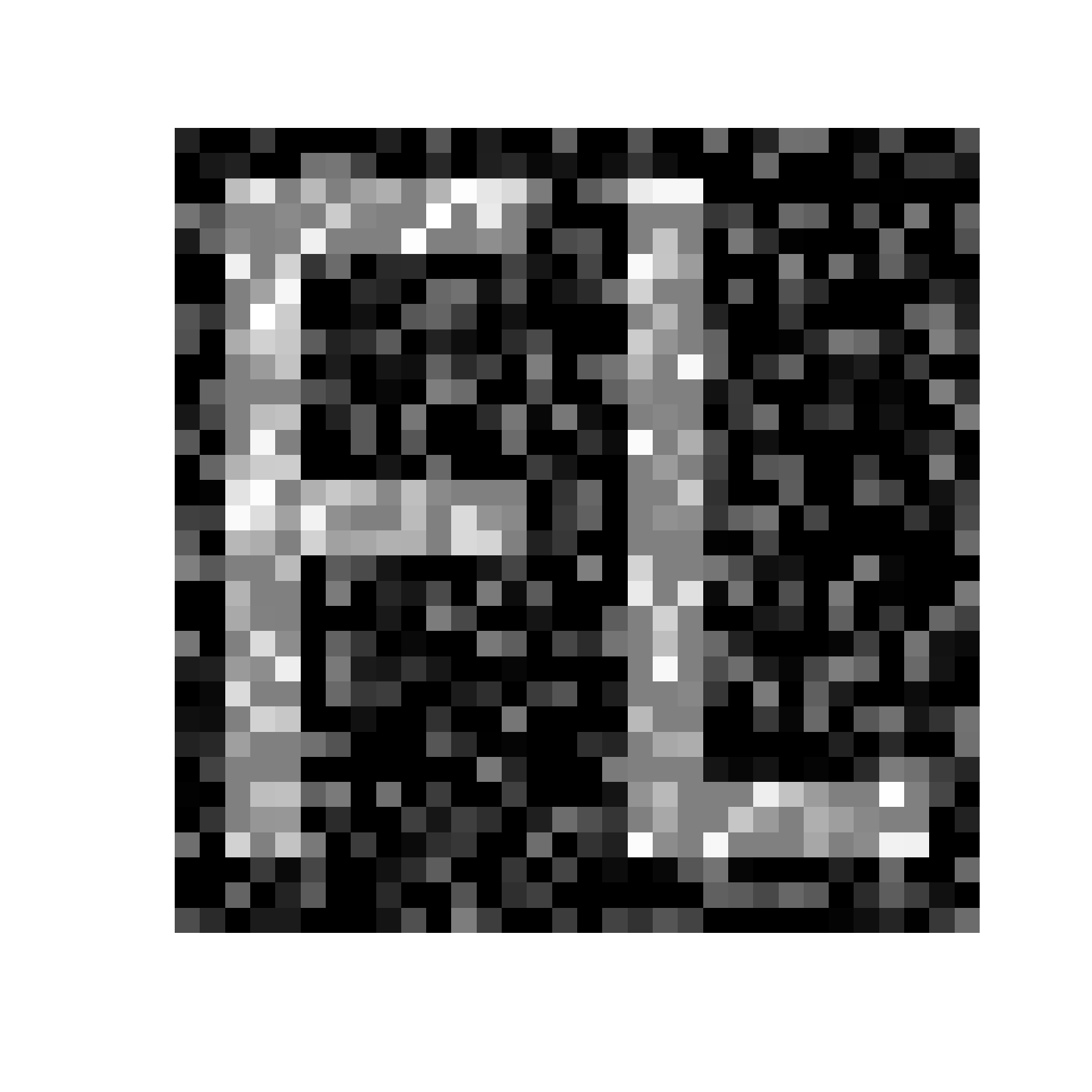}\includegraphics[width=4cm]{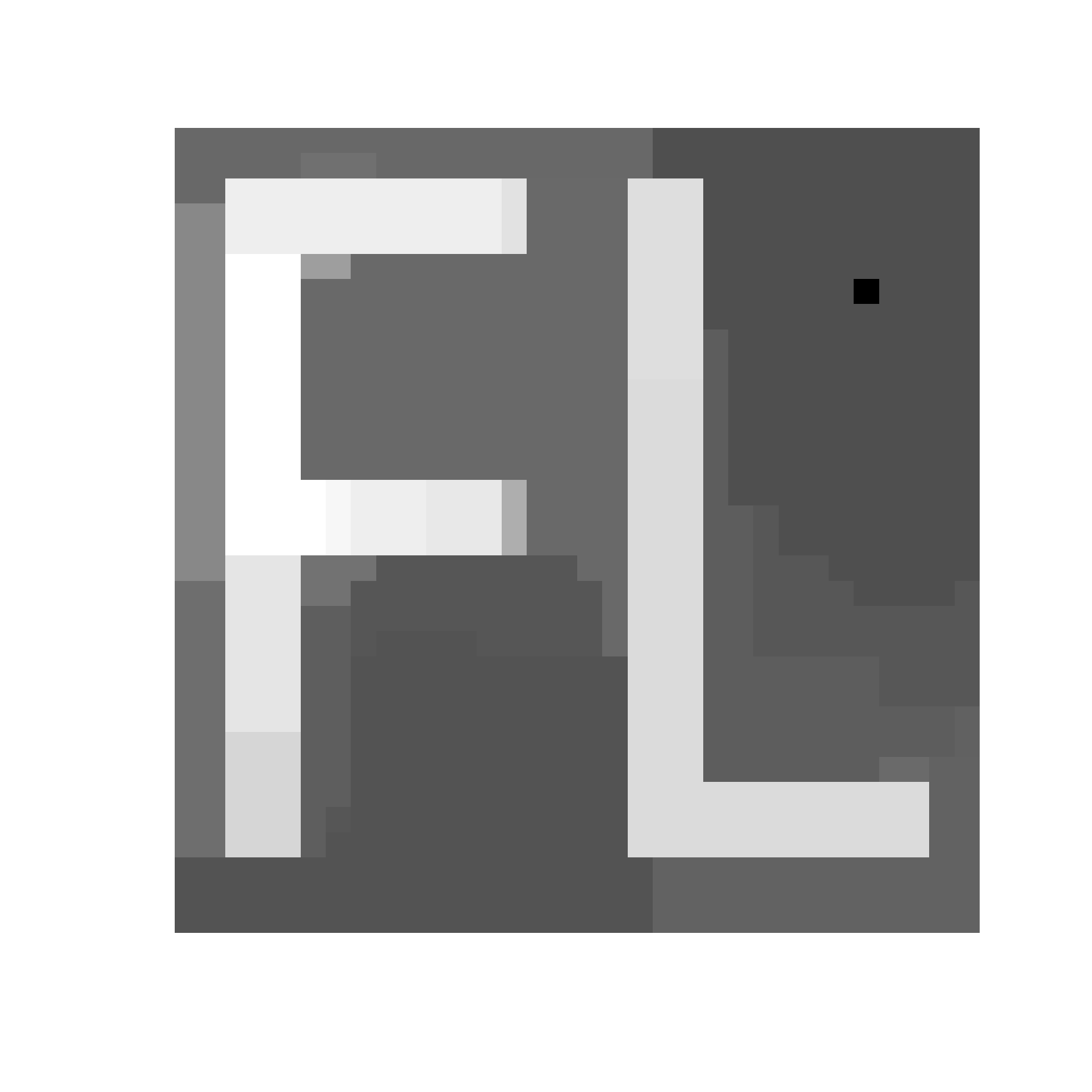}
\caption[Image reconstruction example]{Example of image reconstruction using the Fused Lasso. On the left hand side is the true image. The noisy version is in the middle. The reconstructed version using the Fused Lasso is on the right.\label{fig:FLImage}}
\end{centering}
\end{figure}

Another example where the Fused Lasso model can be used is in image
reconstruction. As a toy example, look at Figure \ref{fig:FLImage}.
On the left hand side, we can see the true image and a noisy
version in the middle. On the right hand side is the denoised version
using the Fused Lasso. As the coefficients are not located on a straight
line but instead on a 2-D grid, we have to use a different
version of the penalty that penalizes all differences of neighboring coefficients in 2 dimensions.

In its more general form, we assume that each coefficient corresponds to a node in a graph $\mathcal{G}=(V,E)$. Then we penalize every difference of coefficients if the corresponding nodes have an edge between them. Specifically, the loss function becomes in this case
\[
\frac{1}{2}\sum_{s}^{n}(y_s-\beta_s)^{2}+\lambda_{1}\sum_{s}^{n}|\beta_{s}|+\lambda_{2} \sum_{(s,t)\in E; s<t}|\beta_s-\beta_t|,
\]
which we will refer to as the general Fused Lasso Signal Approximator
(FLSA). In the example above, the graph is a 2-D grid.

These examples are only special cases of a more general Fused Lasso
model. In this more general form, the Fused Lasso loss function is
\[
L(\by,\bX,\bbeta)=\frac{1}{2}(\by-\bX\bbeta)^{T}(\by-\bX\bbeta)+\lambda_{1}\sum_{i=1}^{p}|\beta_{i}|+\lambda_{2}\sum_{(i,j)\in E,i<j}|\beta_{i}-\beta_{j}|
\]
where as above $E$ are the edges in the graph $\mathcal{G}=(V,E)$ with $V=\{1,\ldots,p\}$
representing the variables. Due to space constraints, the algorithm for the general Fused Lasso for matrices $\bX \in \mathds{R}^{n,p}$ where $\text{rank}(X)=p$ and general graphs $\mathcal{G}$
will be given in the Online Supplement in Section 4. For other matrices with $\text{rank}(X)<p$ we do not develop a path algorithm as the solution path w.r.t. $\lambda_2$ can have discontinuities in this case.
In the following sections, we will present path algorithms for the Fused Lasso Signal Approximator
in its special one-dimensional and its general form. In Section \ref{sec:1DimFLSA}
we will use the special structure of the one-dimensional Fused Lasso
Signal Approximator to derive a fast algorithm that calculates the
entire solution path with a complexity of $n\log n$. The general
FLSA will be treated in Section \ref{sec:GeneralFLSA} and is more
complicated than the one-dimensional case due to the general structure
of the penalty graph $\mathcal{G}$. We prove that our algorithm yields the exact solution and present simulation studies that compare the new algorithms to existing methods. 
Finally, in Section \ref{sec:Conclusion} we will discuss the
results and give possible extensions of these algorithms.

\section{One-dimensional Fused Lasso Signal Approximator\label{sec:1DimFLSA}}

We already used the one-dimensional FLSA in the CGH data example above.
Now, in order to develop a path algorithm, we will first have another
look at the loss function we seek to minimize:
\begin{equation}
L(\by,\bbeta)=\frac{1}{2}\sum_{i=1}^{n}(y_{i}-\beta_{i})^{2}+\lambda_{1}\sum_{i=1}^{n}|\beta_{i}|+\lambda_{2}\sum_{i=1}^{n-1}|\beta_{i}-\beta_{i+1}|.\label{eq:OneDimFLSALoss}
\end{equation}
Due to the simple structure of the loss function, it is possible in
this case to obtain the solution for any value of $(\lambda_{1},\lambda_{2})$
by simple soft-thresholding of the solution obtained for $(0,\lambda_{2}).$
To be more precise, the following theorem holds:

\begin{thm}\label{SoftThresholdingThm}
Assume we have $\bX=\bI$ and that the solution for
$\lambda_{1}=0$ and $\lambda_{2}>0$ is known and denote it by $\bbeta(0,\lambda_{2})$.
Then the solution for $\lambda_{1}>0$ is
\[
\beta_{i}(\lambda_{1},\lambda_{2})=\mbox{sign}\left(\beta_{i}(0,\lambda_{2})\right)\left(|\beta_{i}(0,\lambda_{2})|-\lambda_{1}\right)^{+} \quad \text{for} \quad i=1,\ldots,p.
\]
\end{thm}

The proof of this theorem is presented in \citet{Friedman2007}. It should be noted
that it also holds for the FLSA for arbitrary graphs $\mathcal{G}$,
so that it can also be used for the more general FLSA algorithm. For
the rest of this section, we assume that $\lambda_{1}=0$. The algorithm
presented is a path algorithm that finds the solution for all possible
values of $\lambda_{2}$. Any solution for a $\lambda_{1}\neq0$ can
then be obtained by simply soft-thresholding as shown above.

The path algorithm will start by setting $\lambda_{2}=0$ and then
increase it until all coefficients $\beta_{i}$ have the same value.
For increasing $\lambda_{2}$, neighboring coefficients are forced
to be equal to each other. Once this happens, these coefficients are
being \emph{fused }and subsequently treated as a single variable for
increasing $\lambda_{2}$. In order to be able to do this, it is important
to make sure that these coefficients cannot become \emph{unfused}
again for increasing $\lambda_{2}$. This is in fact the case and
will be shown below. Before getting into the details of the algorithm,
it is necessary to define some notation.

\subsection{Algorithm}

In order to develop the algorithm for the one-dimensional case, we
first have to define what exactly the sets of fused coefficients are. 

\begin{definition}
Let $F_{i}$, $i=1,\ldots,n_{F}(\lambda_{2})$
be the sets of coefficients at $\lambda_{2}$ that are considered
to be fused where $n_{F}(\lambda_{2})$ is the number of such sets.
In order for these sets to be \emph{valid}, every set $F_{i}$ has to be of the form $F_{i}=\{k|l_{i}\le k\le u_{i}\}$
and the following statements have to hold as well:
\begin{itemize}
\item $\cup_{i=1}^{n_{F}(\lambda_{2})}F_{i}=\{1,\ldots,n\}$
\item $F_{i}\cap F_{j}=\emptyset\,\,\, i\neq j$ 
\item Assuming the $F_{i}$ are ordered, for every $k,l\in F_{i}$ we have
$\beta_{k}(\lambda_{2})=\beta_{l}(\lambda_{2})$ and for $k\in F_{i},l\in F_{i+1}$
it holds that $\beta_{k}(\lambda_{2})\neq\beta_{l}(\lambda_{2})$.\label{def:FusedSets1DimFLSA}
\end{itemize}
\end{definition}

For notational convenience, write $\beta_{F_{i}}(\lambda_{2})$ for
any $\beta_{k}(\lambda_{2})$ with $k\in F_{i}$ and also suppress
the dependency of $F_{i}$ onto $\lambda_{2}$. Using this definition
of fused sets, let us now turn to the algorithm.

For the one-dimensional FLSA, a special result holds that makes the
algorithm especially simple and also has been presented in \citet{Friedman2007}.
Loosely speaking, it states that if coefficients are fused at $\lambda_{2}^{0}$,
then these coefficients will also be fused for any $\lambda_{2}>\lambda_{2}^{0}$.
To be more precise

\begin{thm}\label{thm:groupsRemainFused}
Let $\beta_{k}(\lambda_{2})$
be the optimal solution to the one-dimensional FLSA problem for coefficient
$k$ and penalty parameter $\lambda_{2}$. Then if for some $k$ and
$\lambda_{2}^{0}$ it holds that $\beta_{k}(\lambda_{2}^{0})=\beta_{k+1}(\lambda_{2}^{0})$,
then for any $\lambda_{2}>\lambda_{2}^{0}$ it holds that $\beta_{k}(\lambda_{2})=\beta_{k+1}(\lambda_{2})$.
\end{thm}

A proof of this theorem is provided
in \citet{Friedman2007} and an alternative proof using a different
technique is given in the Online Supplement in Section 2.3.

Using this theorem, the algorithm is very simple. First, we need a
starting point for the path algorithm and as the optimal solution
is known for $\lambda_{2}=0$, which is just $\beta_{k}(0)=y_{k}$
for all $k$, we use it to begin the path. Then the algorithm calculates
step by step when two neighboring sets have equal coefficients and
merges them. In order to calculate the slope of the coefficient paths
for some value of $\lambda_{2}$, we assume that we know $\beta_{k}(\lambda_{2})$
for all $k$ as well as the sets of fused variables $F_{i}$. Using
this information, we can calculate the derivative of $\beta_{k}(\lambda_{2})$
with respect to $\lambda_{2}$ and we will see that these are actually
constant, so that the resulting solution path is a piecewise linear
function where the breakpoints occur when two set of coefficients
are being fused. In order to find $\partial\beta_{k}(\lambda_{2})/\partial\lambda_{2}$,
define the loss function $L_{F,\lambda_{2}}$ that incorporates the
fused sets $F_{i}$. This is done by taking the loss function $L$
in Equation (\ref{eq:OneDimFLSALoss}) and replacing $\beta_{k}$ by
$\beta_{F_{i}}$ for $k\in F_{i}$. This constrained loss function
is then
\begin{align*}
L_{F,\lambda_{2}}(\by,\bbeta) & =\frac{1}{2}\sum_{i=1}^{n_{F}(\lambda_{2})}\left(\sum_{j\in F_{i}}(y_{j}-\beta_{F_{i}})^{2}\right)+\lambda_{2}\sum_{i=1}^{n_{F}(\lambda_{2})-1}|\beta_{F_{i}}-\beta_{F_{i+1}}|.
\end{align*}
Due to the assumptions on the sets $F$, the constrained loss function
$L_{F,\lambda_{2}}$ is always differentiable with respect to $\beta_{F_{i}}$
unless two coefficients are just being fused (which only happens if
$\beta_{F_{i}}=\beta_{F_{i+1}}$ for some $i$). Assuming that $\beta_{F_{i}}$
is optimal, the derivative of $L_{F,\lambda_{2}}$ has to be 0 and
we get
\begin{align*}
\frac{\partial L_{F,\lambda_{2}}}{\partial\beta_{F_{i}}} & =|F_{i}|\beta_{F_{i}}-\sum_{j\in F_{i}}y_{j}+\lambda_{2}\; \mbox{sign}(\beta_{F_{i}}-\beta_{F_{i-1}})+ \\
&+ \lambda_2\; \mbox{sign}(\beta_{F_{i}}-\beta_{F_{i+1}})=0\quad \text{for} \quad i=1,\ldots,n_{F}(\lambda_{2})
\end{align*}
where we set $\mbox{sign}(\beta_{F_{1}}-\beta_{F_{0}})=0$ and $\mbox{sign}(\beta_{F_{n_{F}(\lambda_{2})}}-\beta_{F_{n_{F}(\lambda_{2})+1}})=0$.
Now, taking derivatives with respect to $\lambda_{2}$ in these equations
gives the results
\begin{align*}
\frac{\partial\beta_{F_{i}}}{\partial\lambda_{2}}  &=-\frac{1}{|F_{i}|}\left(\mbox{sign}(\beta_{F_{i}}-\beta_{F_{i-1}})+\mbox{sign}(\beta_{F_{i}}-\beta_{F_{i+1}})\right)\quad \text{for} \quad i=1,\ldots,n_{F}(\lambda_{2})
\end{align*}
As we already mentioned above, these are constant as long as the sets
of fused coefficients do not change. By Theorem \ref{thm:groupsRemainFused},
we know that the only way to change the sets of fused coefficients is to merge
two sets as they can never split for increasing $\lambda_{2}$. Therefore,
the solution path is piecewise linear and for increasing $\lambda_{2}$
the breakpoint occurs when two groups fuse. Thus, it is easy to calculate
the next breakpoint, which occurs when neighboring sets have the same
coefficients. In order to do this, define 
\[
h_{i,i+1}(\lambda_{2})=\frac{\beta_{F_{i}}(\lambda_{2})-\beta_{F_{i+1}}(\lambda_{2})}{\frac{\partial\beta_{F_{i+1}}}{\partial\lambda_{2}}-\frac{\partial\beta_{F_{i}}}{\partial\lambda_{2}}}+\lambda_{2} \quad \text{for} \quad i=1,\ldots,n_{F}(\lambda_{2})-1
\]
which is the value for $\lambda_{2}$ at which the coefficients of
the sets $F_{i}$ and $F_{i+1}$ have the same value and can be fused,
assuming that no other coefficients become fused before that. If $h_{i,i+1}(\lambda_{2})<\lambda_{2}$,
these values are being ignored as the two groups $F_{i}$ and $F_{i+1}$
are actually moving apart for increasing $\lambda_{2}$. The next
value at which coefficients are fused is therefore the hitting time
\[
h(\lambda_{2})=\min_{h_{i,i+1}>\lambda_{2}}h_{i,i+1}(\lambda_{2}).
\]
As we are taking the minimum, it is only defined if there is at least
one $h_{i,i+1}>\lambda_{2}$. From equation (\ref{eq:OneDimFLSALoss})
with $\lambda_{1}=0$ we can easily see that for $\lambda_{2}\rightarrow\infty$
the solution is $\beta_{k}=\frac{1}{n}\sum_{l=1}^{n}y_{l}$ for all $k$, thus
only one group exists for large $\lambda_{2}$. Therefore, if $n_{F}(\lambda_{2})\ge2$,
then there exists an $h_{i,i+1}>\lambda_{2}$ and therefore $h(\lambda_{2})$
is defined. Based on these results, we can now write out the details
of the algorithm that provides the entire solution path and they can be found in Algorithm \ref{OneDimFLSAAlg}.

\begin{algorithm}[tbp]
\caption{One-dimensional FLSA path algorithm}
\SetLine
\SetKwBlock{Initialize}{initialize}{end}
\Initialize{$\lambda_{2}=0$\;
$\beta_{k}=y_{k}$ for $k=1,\ldots,n$\;
$F_{i}=\{i\}$ for $i=1,\ldots,n$\;
$n_{F}=n$\;
}%
\While{$n_{F}>1$}{
Calculate next hitting time $h(\lambda_{2})$\;
Let $(i_{0}(\lambda_{2}),i_{0}(\lambda_{2})+1)=\arg\min_{h_{i,i+1}(\lambda_{2})>\lambda_{2}}h_{i,i+1}(\lambda_{2})$
be the indices of the sets to fuse next\;
Fuse the two sets $F_{i_{0}(\lambda_{2})}$ and $F_{i_{0}(\lambda_{2})+1}$\;
Set $\lambda_{2}:=h(\lambda_{2})$\;
Update the values for $\beta_{k}(\lambda_{2})$ , $\frac{\partial\beta_{k}(\lambda_{2})}{\partial\lambda_{2}}$
and set $n_{F}=n_{F}-1$\;
}
\label{OneDimFLSAAlg}
\end{algorithm}

As we will show below, this algorithm only requires a low number of
computational steps and is of complexity $n\log(n)$. Of course, apart
from the computational complexity, it is also important to be able
to save the results in an efficient manner. This can be done with
memory usage on the order $O(n)$. A more detailed analysis can be found in
the Online Supplement in Section 2.

\subsection{Speed comparison}

In order to evaluate the speed of our new algorithm, we want to compare
it to other methods that have been published before. The first alternative
we also use is the component-wise algorithm presented in \citet{Friedman2007}.
The second is based on the general convex solver \emph{CVX}, a package
for specifying and solving convex problems (see \citet{Grant2008a,Grant2008b}).
CVX is very easy to use and flexible, which is why we chose it, despite
the disadvantage that it cannot be used with a warm start. 

\begin{figure}
\begin{centering}
\includegraphics[width=8cm]{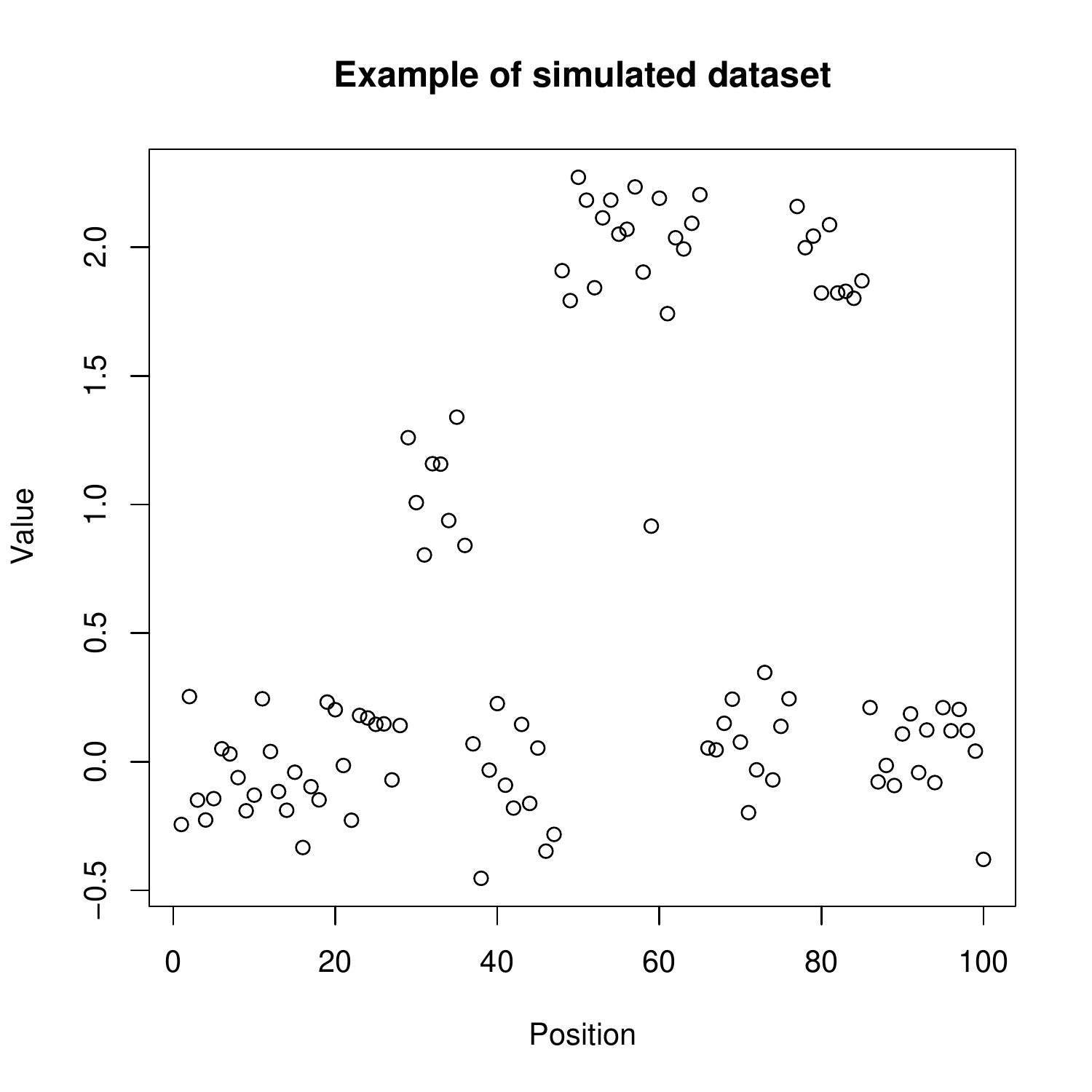}
\caption[Example of 1-dim dataset]{An example of a simulated dataset with size $n=100$ for the one-dimensional
FLSA.}
\label{fig:ExampleSim1Dim}
\end{centering}
\end{figure}

As datasets of a wide range of sizes are needed, the speed comparisons
will be performed on simulated data. The simulated dataset consists
of datapoints with values of 0, 1 and 2. Roughly 20\% of datapoints
will have value 1 and 20\% value 2. An example plot of a simulated
dataset of size $n=100$ can be seen in Figure \ref{fig:ExampleSim1Dim}.

When calculating the solution, our new path algorithm and the competing
methods take somewhat different approaches. Our algorithm calculates
the whole solution path whereas the competing method calculate the
solution only for a prespecified list of $\lambda_{2}$ values. In
order to make the two approaches comparable we measure the time each
algorithm takes to find the solutions for 50 values of $\lambda_{2}$
which are equally spaced between 0 and 1. The results of the comparison
can be found in Table \ref{OneDimFLSATimes}.

As it can be seen, the path algorithm is consistently faster than
the component-wise optimization algorithm for all but the largest
problems. They are also both much faster than the general convex solver
CVX. In addition to this, the path algorithm also returns an object
that stores the complete solution path in a compact form and can be
used to extract solutions for additional values of $\lambda_{2}$
very quickly.

After deriving the path algorithm for the one-dimensional Fused Lasso
Signal Approximator, we want to generalize the algorithm to the case
of the general Fused Lasso Signal Approximator. The most important
difference to the previous algorithm is that a set of fused coefficients
can also break into several sets for increasing values of $\lambda_{2}$.
We will get into more detail in the next section. 

\begin{table}
\begin{centering}
\begin{tabular}{|c|c|c|c|c|c|c|}
\hline 
n & $10^{2}$ & $10^{3}$ & $10^{4}$ & $10^{5}$ & $10^{6}$ & $10^{7}$\tabularnewline
\hline
\hline 
CVX & 17.1 & 30.2 & 210 & 3600 & $>$5 hours & $>$5 hours\tabularnewline
\hline 
Component-wise Alg. & 0.071 & 0.081 & 0.24 & 1.1 & 10 & 98\tabularnewline
\hline 
Path Alg. & 0.0006 & 0.003 & 0.030 & 0.52 & 7.8 & 108\tabularnewline
\hline
\end{tabular}
\end{centering}
\begin{centering}
\caption[Speed comparison for 1-dimensional FLSA]{Time in seconds for a 1-dimensional FLSA problem of size $n$. All
three algorithms calculate the solution for 50 equally spaced values
of $\lambda_{2}$ from $0$ to $1$. Results averaged over 10 simulations.}
\label{OneDimFLSATimes}
\end{centering}
\end{table}

\section{General Fused Lasso Signal Approximator\label{sec:GeneralFLSA}}

In the introduction we have already seen an example where a more general
penalty structure than in the one-dimensional FLSA can be very useful
for reconstructing a noisy image. However, we do not need to restrict
our attention to a two-dimensional grid. In this section, we will
present an algorithm that finds the solution for the FLSA problem
with an arbitrary graph $\mathcal{G}=(V,E)$ (with set of vertices
$V=\left\{ 1,\ldots,n\right\} $ and edges $E$) specifying the structure
of the penalty parameter on differences. The loss function in this
case is
\begin{equation}
L(\by,\bbeta)=\frac{1}{2}\sum_{i=1}^{n}(y_{i}-\beta_{i})^{2}+\lambda_{1}\sum_{i=1}^{p}|\beta_{i}|+\lambda_{2}\sum_{(i,j)\in E,i<j}|\beta_{i}-\beta_{j}|
\label{eq:LossFLSAGeneral}
\end{equation}
so that we penalize $|\beta_{i}-\beta_{j}|$ for every edge $(i,j)\in E$.
The condition $i<j$ only makes sure that we penalize $|\beta_{i}-\beta_{j}|$
only once as the edges in the graph are assumed to be undirected.
As in the previous section, we can use the soft-thresholding theorem
(see Theorem \ref{SoftThresholdingThm}) and it is therefore possible
to set $\lambda_{1}=0$ and find a solution path for $\lambda_{2}$
and later obtain any solution for $\lambda_{1}>0$ by soft-thresholding.
The algorithm for this more general case is conceptually similar to
the one presented for the one-dimensional case. However, unlike in
this simpler setup, for the more general penalty structure it is not
guaranteed that a group of coefficients that is fused for value
$\lambda_{2}=\lambda_{2}^{0}$ will remain fused for $\lambda_{2}>\lambda_{2}^{0}$,
but instead fused groups may break up for increasing $\lambda_{2}$.
The main adjustment to deal with this problem will be to introduce
a method to determine for which value of $\lambda_{2}$ a group of
fused variables will break up. 

In the following subsections, we will first make some necessary adjustments
to the definition of sets of fused variables and given these sets,
we calculate $\partial\beta_{F_{i}}/\partial\lambda_{2}$.
Next, we give the conditions under which a set of fused coefficients
breaks up into two smaller sets and present a method on how to calculate
critical values of $\lambda_{2}$ for which this could happen. After
incorporating this into the final algorithm, we present an approximate
version of our method that is faster on large dataset by sacrificing some precision. 
Finally, we use our new algorithm on simulated
data and compare its speed and accuracy to the other methods we have
already used for the one-dimensional FLSA. Now, let us first make
the necessary changes to the definition of fused coefficients to account
for the general graph structure $\mathcal{G}$.

\subsection{Sets of fused variables }

In the case of the one-dimensional FLSA before, we already specified
certain conditions for the sets of fused variables $F_{i}$ in Definition
\ref{def:FusedSets1DimFLSA}. Here, due to the more complicated structure
of the graph $\mathcal{G}$, we have to restate the condition that
any set has to be an interval. For general graphs, the condition is
instead that any set of fused variables has to be connected. For the
definition and also in the following sections, assume that we know
the minimizer of the loss function for penalty parameter $\lambda_{2}$
and denote it by $\beta_{F_{i}}(\lambda_{2})$. Then the definition
of a valid set of fused variables is:

\begin{definition}Let $n_{F}(\lambda_{2}^{0})$ be the number of
sets of fused variables for penalty parameter $\lambda_{2}^{0}$.
Then for the sets $F_{i}$, $i=1,\ldots,n_{F}(\lambda_{2}^{0})$ to
be \emph{valid}, the following conditions have to hold:
\begin{enumerate}
\item $\cup_{i=1}^{n_{F}(\lambda_{2})}F_{i}=\{1,\ldots,n\}$
\item $F_{i}\cap F_{j}=\emptyset\,\,\, i\neq j$ 
\item If $k,l\in F_{i}$ then $\beta_{k}(\lambda_{2}^0)=\beta_{l}(\lambda_{2}^0)$
and if $k\in F_{i,}l\in F_{j},i\neq j$ and $F_{i}$ and $F_{j}$
have a connecting edge, then $\beta_{k}(\lambda_{2})\neq\beta_{l}(\lambda_{2})$
for all penalty parameters $\lambda_{2}\in(\lambda_{2}^{0},\lambda_{2}^{0}+\varepsilon)$
for some $\varepsilon>0$.
\item If $k,l\in F_{i}$ then $k$ and $l$ are connected in $\mathcal{G}$
by only going over nodes in $F_{i}$, i.e. $k,l$ are connected in
$\mathcal{G}|_{F_{i}}$, the subgraph of $\mathcal{G}$ induced by
$F_{i}$.
\end{enumerate}\label{def:FusedSetsGeneral}
\end{definition}

Compared to the previous version of this definition, the third and
fourth condition have been adapted. The third condition now reflects
that sets can be split up for increasing $\lambda_{2}$ and the last
one is equivalent to the requirement that $F_{i}$ is an interval
in the one-dimensional FLSA case. 

Now, assuming that the sets of variables $F_{i}$ that are fused  are
given, we will determine the slope of the optimal $\beta_{F_{i}}$
with respect to $\lambda_{2}$. In order to do this, we will incorporate
the sets of fused coefficients into the loss function (\ref{eq:LossFLSAGeneral}).
Setting $\beta_{k}=\beta_{F_{i}}$ for all $k\in F_{i}$, the loss
function becomes
\begin{align}
L_{F,\lambda_{2}}(\by,\bbeta)&=\sum_{i=1}^{n_{F}(\lambda_{2})}\left(\sum_{j\in F_{i}}(y_{j}-\beta_{F_{i}})^{2}\right)+ \notag \\
&+\lambda_{2}\sum_{i<j}\left|\left\{ (k,l)\in E:k\in F_{i},l\in F_{j}\right\} \right|\left|\beta_{F_{i}}-\beta_{F_{j}}\right|.\label{eq:ConstrainedLossFLSAGeneral}
\end{align}

Here, note that by definition of the sets $F_{i}$, we have that $\beta_{F_{i}}(\lambda_{2})\neq\beta_{F_{j}}(\lambda_{2})$
for $j\neq i$ (except for a finite number of $\lambda_{2}$ for which
sets are fused or split) and therefore the loss function is differentiable
with respect to $\beta_{F_{i}}$ at the solution $\beta_{F_{i}}(\lambda_{2})$.
Thus, at $\beta_{F_{i}}(\lambda_{2})$, the derivative of $L_{F,\lambda_{2}}(\by,\bbeta)$
with respect to $\beta_{F_{i}}$ is 0, that is
\begin{align*}
\frac{\partial L_{F,\lambda_{2}}(\by,\bbeta)}{\partial\beta_{F_{i}}}&=|F_{i}|\beta_{F_{i}}-\sum_{j\in F_{i}}y_{j}+\\
&+\lambda_{2}\sum_{j\neq i}\left|\left\{ (k,l)\in E:k\in F_{i},l\in F_{j}\right\} \right|\mbox{sign}\left(\beta_{F_{i}}-\beta_{F_{j}}\right)=0
\end{align*}
for the solutions $\beta_{F_{i}}(\lambda_{2})$. By taking the derivative
w.r.t. $\lambda_{2}$ and noting that for small changes of $\lambda_{2}$,
the sign of $\beta_{F_{i}}-\beta_{F_{j}}$ does not change, it is
possible to determine $\partial\beta_{F_{i}}(\lambda_{2})/\partial\lambda_{2}$
as
\begin{equation}
\frac{\partial\beta_{F_{i}}(\lambda_{2})}{\partial\lambda_{2}}=-\frac{\sum_{j\neq i}\left|\left\{ (k,l)\in E:k\in F_{i},l\in F_{j}\right\} \right|\mbox{sign}\left(\beta_{F_{i}}-\beta_{F_{j}}\right)}{\left|F_{i}\right|},\label{eq:BetaDeriv}
\end{equation}
which is constant as long as the $F_{i}$ do not change. Therefore,
the solution $\beta_{F_{i}}(\lambda_{2})$ is a piecewise linear function
again. At the breakpoints of the solution path, the sets of fused
variables change. As we will see in more detail later, there are 2
things that can happen:

\begin{itemize}
\item $\beta_{F_{i}}(\lambda_{2})=\beta_{F_{j}}(\lambda_{2})$ for some
$i\neq j$ with $F_{i}$ and $F_{j}$ connected, which violates condition
3 of Definition \ref{def:FusedSetsGeneral}. In this case, fuse sets $F_{i}$ and $F_{j}$.
\item A set $F_{i}$ has to be broken up into 2 smaller subsets. A way to
determine when this has to happen is presented below.
\end{itemize}
Once the sets have been updated, the solution path is again linear
so that the whole solution path for $\lambda_{2}$ can be obtained
by updating the sets of fused coefficients at the right values of
$\lambda_{2}$. Now we will take a closer look at the conditions under
which we split a set of fused variables.

\subsection{Splitting and fusing sets of variables}

In order to see when it is necessary to split sets of variables, we
have to see when the solution $\beta_{F_{i}}(\lambda_{2})$ and its
derivatives obtained from the constrained loss function $L_{F,\lambda_{2}}$
from Equation (\ref{eq:ConstrainedLossFLSAGeneral}) is also optimal
for the unconstrained loss $L_{\lambda_{2}}$ in Equation (\ref{eq:LossFLSAGeneral}).
For this, we will look at the subgradient equations of $L_{\lambda_{2}}$.
An overview of subgradients can be found in \citet{Bertsekas1999}.

\subsubsection{Subgradient equations}

As $L_{\lambda_{2}}$ is not differentiable everywhere, it is convenient
to use subgradients instead of the usual derivatives. For the subgradients,
a necessary and sufficient condition for $\beta_{k}$ to be optimal
is that \begin{equation}
\frac{\partial L_{\lambda_{2}}(\by,\bbeta)}{\partial\beta_{k}}=\beta_{k}-y_{k}+\lambda_{2}\sum_{(k,l)\in E}t_{kl}=0\,\,\,\,\,\mbox{for}\,\,\,\,\, k=1,\ldots,n\label{eq:subgradients}\end{equation}
where $t_{kl}=\mbox{sign}(\beta_{k}-\beta_{l})$ for $\beta_{k}\neq\beta_{l}$,
$t_{kl}\in[-1,1]$ for $\beta_{k}=\beta_{l}$. For the case $\beta_{k}=\beta_{l}$ it is also enforced that $t_{kl}=-t_{lk}$ (which is trivially true in the $\beta_k \neq \beta_l$ case).
Given a grouping $F_{i}$, these equations can be written slightly
differently. With $k\in F_{i}$, let
\[
\frac{\partial L_{\lambda_{2}}(\by,\bbeta)}{\partial\beta_{k}}=\beta_{k}-y_{k}+\lambda_{2}\sum_{i\neq j}\sum_{(k,l)\in E:l\in F_{j}}t_{kl}+\lambda_{2}\sum_{(k,l)\in E:l\in F_{i}}t_{kl}=0\,\,\,\,\,\mbox{for}\,\,\,\,\, k=1,\ldots,n
\]
where we grouped the $t_{kl}$ for which $k,l\in F_{i}$. Assuming
that $\beta_{F_{i}}(\lambda_{2})$ is a minimizer of the unconstrained
loss function, there exist $t_{kl}(\lambda_{2})$ such that the subgradient
equations hold. Writing $\tau_{kl}(\lambda_{2})=\lambda_{2}t_{kl}(\lambda_{2})$
and taking the derivative w.r.t. $\lambda_{2}$ in the subgradient
equations, we get
\begin{equation}
\frac{\partial\beta_{k}}{\partial\lambda_{2}}+\sum_{i\neq j}\sum_{(k,l)\in E:l\in F_{j}}t_{kl}+\sum_{(k,l)\in E:l\in F_{i}}\frac{\partial\tau_{kl}}{\partial\lambda_{2}}=0\,\,\,\,\,\mbox{for}\,\,\,\,\, k=1,\ldots,n.\label{eq:DerivSubGrad}
\end{equation}
where we exploit the fact that $t_{kl}=\pm1$ is constant for $k\in F_{i}$
and $l\in F_{j}$. Note that there are not necessarily unique values
for $\partial\tau_{kl}/\partial\lambda_{2}$ such that these
equations hold and there may be an infinite number of possible solutions,
any of which will serve our purpose. Also note that from above we
know that as long as the set $F_{i}$ stays fixed, $\partial\beta_{k}/\partial\lambda_{2}$
is constant for $k \in F_i$. This is the case because $t_{kl}$ for $k \in F_j$ and $l \in F_i$ with $i \neq j$ is defined as $t_{kl}= \sign(\beta_k-\beta_l)$ and stays constant as long as the order of $\beta_k$ and $\beta_l$ does not change. However, if the order changes, then by Definition \ref{def:FusedSetsGeneral} the set $F_i$ has to change as well. Therefore, the equation above stays the same as long as $F_i$ is fixed. Therefore, the $\partial\tau_{kl}/\partial\lambda_{2}$ only
depend on the groups and not on $\lambda_{2}$. However, apart from
Equation (\ref{eq:DerivSubGrad}), it also has to hold that $\tau_{kl}(\lambda_{2})\in[-\lambda_{2},\lambda_{2}]$,
as the condition $t_{kl}\in[-1,1]$ has to hold. Thus, when we keep
the groups $F_{i}$ fixed, it will not always be possible to find
$\partial\tau_{kl}/\partial\lambda_{2}$ such that for increasing
$\lambda_{2}$ these conditions still hold. As $\partial\tau_{kl}/\partial\lambda_{2}$
is constant, the $\tau_{kl}$ are piecewise linear and the $t_{kl}$
are continuous in $\lambda_{2}$. Then there are two ways in which
the subgradient conditions can fail:

\begin{enumerate}
\item There is a group $F_{i}$ for which for increasing $\lambda_{2}$,
the condition $-1\le t_{kl}\le1$ for all $k,l\in F_{i}$ cannot be
satisfied anymore. In this case, the group has to be split into two
smaller subgroups. How to decide when this is the case and how to
identify the two new subgroups will be treated below.
\item There are two groups $F_{i}$ and $F_{j}$ for which there exists
$k\in F_{i}$ and $l\in F_{j}$ with $(k,l)\in E$, i.e. $F_{i}$
and $F_{j}$ have at least one edge connecting them. For these two
groups, we have that $t_{kl}=\pm1$ does not hold anymore for all
$k\in F_{i}$ and $l\in F_{j}$. In this case, the groups $F_{i}$
and $F_{j}$ have to be fused. 
\end{enumerate}
It is very easy to detect when the second case above occurs, i.e.,
when two groups have to be fused. If we have two groups $F_{i}$ and
$F_{j}$ that are directly connected to each other by an edge, then
they will be fused at $\lambda_{2}^{0}$ if $\beta_{F_{i}}(\lambda_{2}^{0})=\beta_{F_{j}}(\lambda_{2}^{0})$
and $\beta_{F_{i}}(\lambda_{2})\neq\beta_{F_{j}}(\lambda_{2})$ for
some interval $\lambda_{2}\in(\lambda_{2}^{0}-\varepsilon,\lambda_{2}^{0})$,
i.e. groups $F_{i}$ and $F_{j}$ {}``hit'' each other at $\lambda_{2}^{0}$.
Detecting when a group has to be broken into two smaller groups is
harder and we will discuss it in the following subsection.

\subsubsection{\label{sub:The-maximum-flow}The maximum flow problem}

In order to decide when to split variables, it is necessary to find
solutions for $t_{kl}(\lambda_{2})$ (or equivalently $\tau_{kl}$)
in Equation (\ref{eq:DerivSubGrad}). For $k$ and $l$ that are not
in the same group, this will be easy as then $t_{kl}=\mbox{sign}(\beta_{F_{i}}-\beta_{F_{j}})$
for $k\in F_{i}$ and $l\in F_{j}$. For $k$ and $l$ in the same
group, i.e. $k,l\in F_{i}$, we will see that $\tau_{kl}$ is an affine
function of $\lambda_{2}$ and the slope can be calculated by solving
a maximum flow problem in graph theory. The same calculations can
also be used to identify when a group has to be split by (more on
this later). As maximum flow problems are a well studied area, fast
algorithms for this problem such as the push-relabel algorithm, among
others, exist and can be used here (see \citet{Cormen2001}). Before
going into more details, assume that for penalty parameter $\lambda_{2}$,
we know $\tau_{kl}(\lambda_{2})$ for $k,l\in F_{i},i=1,\ldots,n_{F}(\lambda_{2})$
that solve the subgradient Equation (\ref{eq:DerivSubGrad}). Now,
for notational convenience, define

\[
p_{k}=-\sum_{i\neq j}\sum_{(k,l)\in E:l\in F_{j}}t_{kl}-\frac{\partial\beta_{F_{i}}(\lambda_{2})}{\partial\lambda_{2}}
\]
for $k\in F_{i}$ and call it the \emph{push} on node $k$ as it measures
the influence other variables in neighboring groups that are connected
to $k$ have on node $k$. Using these definitions, Equation (\ref{eq:DerivSubGrad})
can be written as

\begin{equation}
\sum_{(k,l)\in E:l\in F_{i}}\frac{\partial\tau_{kl}}{\partial\lambda_{2}}=p_{k}\,\,\,\mbox{\,\, for}\,\,\,\,\, k=1,\ldots,n.\label{eq:FlowPush}\end{equation}
For each of these equations, we can see that they only involve variables
that belong the same groups, i.e. if $k\in F_{i}$, then all $l$
used for the variables $\tau_{kl}$ are also in $F_{i}$ and thus
these $n$ equations are separable according to the groups $F_{i}$.
Therefore, for each of the groups $F_{i}$, we will solve a separate
maximum flow problem to find $\partial\tau_{kl}/\partial\lambda_{2}$
and determine if it is necessary to split the group. 

For any maximum flow problem, we need to specify the underlying graph
which consists of vertices, edges and capacities on each edge in both
directions. In order to do this, we first define the graph $\mathcal{G}_{i}=\mathcal{G}|_{F_{i}}$,
which is the graph $\mathcal{G}$ restricted to the nodes in the set
$F_{i}$. Now let $\tilde{\mathcal{G}}_{i}=(\tilde{V_{i},}\tilde{E_{i}},\tilde{C_{i})}$ be
the vertices, edges and capacities of the $i$-th problem. For these,
we define:

\begin{description}
\item [{Vertices:}] To each of the subgraphs $\mathcal{G}_{i}$, we add
an artificial source node $r$ and sink node $s$, such that $\tilde{V}_{i}=V_{i}\cup\{r,s\}$. 
\item [{Edges:}] For the edges $\tilde{E}_{i}$ we use all the edges in
$E_{i}$ and will add additional edges connecting each of the nodes
in $V_{i}$ to either the source or the sink. In order to motivate
which nodes will be connected to which, note that at the end, we will
set $\partial\tau_{kl}/\partial\lambda_{2}=f_{kl}$ for the
maximal flow $f_{kl}$ from node $k$ to node $l$. As the flow through
every node (except for the source and sink) has to be 0, a node has
an edge with the source if the RHS of equation (\ref{eq:FlowPush})
is greater than 0 and an edge with the sink if it is less than 0.
Thus \[
\tilde{E}_{i}=E_{i}\cup\left\{ (r,l):p_{l}>0\right\} \cup\left\{ (k,s):p_{k}<0\right\} \]
As usual, all the edges are undirected. 
\item [{Capacities:}] Of course, the capacities on the edges have to be
defined as well and we will define them for each direction separately.
As we will set $\partial\tau_{kl}/\partial\lambda_{2}=f_{kl}$
for $k,l\in F_{i}$, the flows have to be constrained such that $\tau_{kl}(\lambda_{2})$
stays within the interval $[-\lambda_{2},\lambda_{2}]$. This has
to hold for all edges in $E_{i}$. For the edges to the source or
the sink, the corresponding absolute value on the RHS of Equation
(\ref{eq:FlowPush}) will be used. This way, with $\partial\tau_{kl}/\partial\lambda_{2}=f_{kl}$,
Equation (\ref{eq:FlowPush}) will hold if and only if all edges coming
from the source and all edges going to the sink are at full capacity.
\\
In order to ensure that $\tau_{kl}(\lambda_{2})\in[-\lambda_{2},\lambda_{2}]$,
we have the restrictions\[
\frac{\partial\tau_{kl}}{\partial \lambda_{2}}\in\begin{cases}
(-\infty,\infty) & \mbox{if }\tau_{kl}\in(-\lambda_{2},\lambda_{2})\\
(-\infty,1] & \mbox{if }\tau_{kl}=\lambda_{2}\\
{}[-1,\infty) & \mbox{if }\tau_{kl}=-\lambda_{2}.\end{cases}\]
Correspondingly, we set for $k,l\in F_{i}$\begin{eqnarray*}
(c_{kl},c_{lk}) & = & \begin{cases}
(-\infty,\infty) & \mbox{if }\tau_{kl}\in(-\lambda_{2},\lambda_{2})\\
(1,\infty) & \mbox{if }\tau_{kl}=\lambda_{2}\\
(\infty,1) & \mbox{if }\tau_{kl}=-\lambda_{2}.\end{cases}\end{eqnarray*}
Now it only remains to define the capacities on the edges coming from
the source or going to the sink. First, for the edges from the source
$r$, i.e. all $l$ for which $(r,l)\in\tilde{E}_{i}$,\[
(c_{rl},c_{lr})=\left(p_{l},0\right)\]
and correspondingly for the edges to the sink set for all $k$ with
$(k,s)\in\tilde{E}_{i}$,\[
(c_{ks},c_{sk})=\left(-p_{k},0\right).\]
Using all this set $\tilde{C}_{i}=\left\{ c_{kl}:k,l\in\tilde{V}_{i}\right\} $.
\end{description}
Here it is interesting to note that 

\begin{equation}
\sum_{k\in F_{i}}p_{k}=0\label{eq:sumOfPushs}
\end{equation}
which is easy to see by summing up Eequations (\ref{eq:DerivSubGrad}),
the definition of $p_{k}$ and that $t_{kl}=-t_{lk}$. Therefore,
the sum of all capacities going out of the source is equal to the
sum of all capacities going into the sink and thus a flow that is
maximal for all source edges is also maximal for all sink edges. 

Now that we have defined the maximal flow problem, we will not go
into any detail of how to solve it and just refer to the literature
that we have already mentioned above. In the following, the solution
to the flow problem will be referred to as $f_{kl}$, which is the
flow from node $k$ to node $l$, assuming that $k,l\in\tilde{E}_{i}$.
Using this result we will now show that the solution path is piecewise-linear.
The next theorem guarantees that for an interval, the solution for
$\beta_{k}(\lambda_{2})$ and $\tau_{kl}(\lambda_{2})$ are affine
and have the slope as stated above. 

\begin{thm}\label{ThmSolutionLinear}
For some $\lambda_{2}^{0}$,
let $F_{1},\ldots,F_{n_{F}(\lambda_{2}^{0})}$ be a valid grouping
of the variables. Let $\tilde{\mathcal{G}_{i}}$ be the with $F_{i}$
associated maximum-flow graph as defined above. Also let $\beta_{k}(\lambda_{2}^{0})$
and $\tau_{kl}(\lambda_{2}^{0})$ be a solution to the FLSA problem
for penalty parameter $\lambda_{2}=\lambda_{2}^{0}$. If $\tilde{\mathcal{G}_{i}}$
has a maximum flow for which all flows coming from the source are
at maximum capacity (i.e. $f_{rl}=c_{rl}$ for all $(r,l)\in\tilde{E}_{i}$),
and $\frac{\partial\beta_{F_{i}}}{\partial\lambda_{2}}$ is as defined
in equation (\ref{eq:BetaDeriv}), then there exists some $\Delta>0$
such that for any $\lambda_{2}\in[\lambda_{2}^{0},\lambda_{2}^{0}+\Delta]$,
the solution to the FLSA problem is given by
\[
\beta_{k}(\lambda_{2})=\beta_{k}(\lambda_{2}^{0})+\frac{\partial\beta_{F_{i}}}{\partial\lambda_{2}}(\lambda_2^0) \cdot (\lambda_{2}-\lambda_{2}^{0})\quad\mbox{for}\quad k\in F_{i}
\]
and
\begin{eqnarray*}
\tau_{kl}(\lambda_{2}) & = & \begin{cases}
\tau_{kl}(\lambda_{2}^{0})+f_{kl}(\lambda_{2}-\lambda_{2}^{0}) & \mbox{for }k,l\in F_{i}\quad\mbox{for some }i\\
\mbox{sign}(\tau_{kl}(\lambda_{2}^{0}))\lambda_{2} & \mbox{otherwise}.\end{cases}
\end{eqnarray*}
\end{thm}

The proof of this theorem can be found in the Online Supplement in Section 3. The only item
in the previous proof that we haven't specified so far is the length
of the interval $\Delta$, for which the solution will be linear as
described. There are two things that can occur, that would violate
the assumptions of Theorem \ref{ThmSolutionLinear}. First, two sets
of variables $F_{i}$ and $F_{j}$ that have a connecting edge have
$\beta_{F_{i}}=\beta_{F_{j}}$ and therefore, the grouping is not
valid anymore. In this case, the two sets have to be merged. Second,
for some group $F_{i}$, the maximum flow problem is not at maximal
capacity for all source nodes. Then, this group has to be split. However,
before proving that these operations yield a valid grouping for which
the assumption of Theorem \ref{ThmSolutionLinear} hold, we will determine
$\Delta$.

\subsubsection{The hitting time $h$ and splitting time $v$}

In order to find $\Delta,$ we will first determine the smallest value
$\lambda_{2}>\lambda_{2}^{0}$ where two neighboring groups have the
same coefficient. Next, we will determine the smallest $\lambda_{2}>\lambda_{2}^{0}$
such that the conditions on $\tau_{kl}$ are violated. $\lambda_{2}^{0}+\Delta$
is then the smallest of these two values. 

Start by assuming that for penalty parameter $\lambda_{2}^{0}$ we
have a valid grouping $F_{1},\ldots,F_{n_{F}(\lambda_{2}^{0})}$ and
solutions $\beta_{k}(\lambda_{2}^{0})$ as well as $\tau_{kl}(\lambda_{2}^{0})$.
Given this, it is easy to calculate when two sets that are connected
by an edge hit. For this, let the hitting time of groups $i$ and
$j$ at $\lambda_{2}^{0}$ be\[
h_{ij}(\lambda_{2}^{0})=\begin{cases}
(\beta_{F_{i}}-\beta_{F_{j}})/\left(\frac{\partial\beta_{F_{j}}}{\partial\lambda_{2}}-\frac{\partial\beta_{F_{i}}}{\partial\lambda_{2}}\right)+\lambda_{2}^{0} & \mbox{if }\exists k\in F_{i},l\in F_{j}\mbox{ with }(k,l)\in E\\
\infty & \mbox{otherwise.}\end{cases}\]
If $h_{ij}<\lambda_{2}^{0}$, then given the current slopes, these
groups will not meet for $\lambda_{2}>\lambda_{2}^{0}$. If $h_{ij}=\lambda_{2}^{0}$,
then $\beta_{F_{i}}(\lambda_{2}^{0})=\beta_{F_{j}}(\lambda_{2}^{0})$.
However, as we assumed that this is a valid grouping, from the definition
we get that for some $\varepsilon>0$, for any $\lambda_{2}\in(\lambda_{2}^{0},\lambda_{2}^{0}+\varepsilon)$
we have $\beta_{F_{i}}(\lambda_{2})\neq\beta_{F_{j}}(\lambda_{2})$
and as the trajectories are piecewise affine, the groups $F_{i}$
and $F_{j}$ move apart. Therefore, defining
\[
h(\lambda_{2}^{0})=\min_{h_{ij}>\lambda_{2}^{0}}h_{ij}(\lambda_{2}^{0})
\]
we have that two groups will hit at $h$ but not before. Therefore,
the grouping remains valid for at least an interval $h-\lambda_2^0$. 

Now let us look at the maximum flow problem and how long the interval
can be such that for all $\tau_{kl}(\lambda_{2})\in[-\lambda_{2},\lambda_{2}]$.
Then, given the flows $f_{kl}$, define the violation time of the constraint on $\tau_{kl}$ as 
\[
v_{kl}(\lambda_{2}^{0})=\begin{cases}
\frac{|\mbox{sign}(f_{kl})\lambda_{2}^{0}-\tau_{kl}(\lambda_{2}^{0})|}{|f_{kl}|-1}+\lambda_{2}^{0} & \mbox{if }|f_{kl}|>1\\
\infty & \mbox{otherwise,}\end{cases}
\]
and set
\[
v(\lambda_2^0)=\min v_{kl}(\lambda_2^0).
\]
Then given the behavior of the $\tau_{kl}$ described in Theorem \ref{ThmSolutionLinear},
we have that $\tau_{kl}(\lambda_{2})\in[-\lambda_{2},\lambda_{2}]$
for $\lambda_{2}\in[\lambda_{2}^{0},v]$. Furthermore, for $\lambda_{2}>\lambda_{2}^{0}+v$,
at least one of the constraints would be violated. However, note that
this does not necessarily mean that the group has to be split. It
may just be necessary to find a new maximal flow that satisfies the
constraints. Using the hitting and violation time, we can set 
\[
\Delta=\min(h(\lambda_2^0),v(\lambda_2^0))-\lambda_{2}^{0}.
\]
We can now distinguish two cases:
\begin{description}
\item [Case 1 {$h\le v$}:] Here, the two sets that hit at $\lambda_{2}=h$ have
to be merged.
\item [Case 2 {$h>v$}:] In one of the sets, say $F_{i}$, for at least one edge
$(k,l)$ with $k,l\in F_{i}$ we have $\tau_{kl}(v)=\pm v$ and if
the slope remained unchanged, then $|\tau_{kl}(\lambda_{2})|>|\lambda_{2}|$
for $\lambda_{2}>v$, thus the constraint would be violated. Then
the capacity constraints of the associated maximum flow problem has
to be updated and a new maximum flow identified. If for the new flow,
all source edges are at capacity, only the trajectories for the $\tau_{kl}$
have to be altered. Otherwise, the set has to be split.
\end{description}
Therefore, we have defined the $\Delta$ of the previous theorem and
identified the values of $\lambda_{2}$ at which the piecewise-linear
solution path has breakpoints.

\subsubsection{Adapting the sets of fused variables\label{sub:Adapting-the-sets}}

Of course, we still have to specify how to exactly split a set for
which the source edges are not at capacity into two smaller subsets.
Assume that $F_{i}$ is the set that has to be split with associated
maximum flow graph $\tilde{\mathcal{G}}_{i}$. Then define the set
\[
R_{i}=\left\{ l\in F_{i}:r\mbox{ connected to }l\mbox{ by an augmenting path in }\tilde{\mathcal{G}}_{i}\right\} 
\]
where the augmenting path is defined with respect to the maximal flow
$f_{kl}$, i.e. for each node $l\in R_{i}$ there exists a path from
the source $r$ to $l$ using only edges for which the flow is not
at capacity. The complement of $R_{i}$ with respect to $F_{i}$ is
defined as $S_{i}=F_{i}\backslash R_{i}$. Then we divide the set
$F_{i}$ into the two subsets $R_{i}$ and $S_{i}$. 

Now it remains to be shown that fusing or splitting sets as described
above will yield sets of fused variables that satisfy the assumptions
of Theorem \ref{ThmSolutionLinear}. In particular, whenever we are
at a breakpoint, i.e. have to fuse or split sets or both, we propose
the following procedure for adapting the sets of fused variables:

\begin{enumerate}
\item If there are sets $F_{i}$ and $F_{j}$ for which $\exists k\in F_{i}$
and $l\in F_{j}$ with $(k,l)\in E$ and $\beta_{F_{i}}=\beta_{F_{j}}$,
then fuse these sets into a new set $\tilde{F}_{ij}=F_{i}\cup F_{j}$
if $(\partial\beta_{F_{i}}/\partial\lambda_{2})-(\partial\beta_{F_{j}}/\partial\lambda_{2})\le0$
and $t_{kl}=1$.
\item If there is a set $F_{i}$ for which in the associated maximal flow
graph not all edges coming from the source are at maximal capacity,
then split $F_{i}$ in the two subsets $R_{i}$ and $S_{i}$ as described
above. 
\item Iterate steps 2 and 3 until nothing changes. 
\end{enumerate}
Using this procedure we can now show that adapting the sets in this
way is correct.

\begin{prop}\label{PropFuseSplitSets}Assume that we perform the
fusion and split steps as described above. Then, the algorithm stops
after a finite number of fusion and splitting steps and the resulting
sets of variables are valid and satisfy the assumptions of Theorem
\ref{ThmSolutionLinear}. \end{prop}

Again, the proof can be found in the Online Supplement in Section 3. Putting all this together,
we have shown that the solutions are piecewise linear and how to change
the sets of fused variables at the breakpoints. So overall, using
this algorithm we can calculate the entire solution path.

\subsection{Outline of the algorithm}

In the previous sections, we have seen how to derive the entire solution
path of the general Fused Lasso Signal Approximator. Overall, the
algorithm is very similar to the one-dimensional FLSA outlined in
Algorithm \ref{OneDimFLSAAlg}. The most important change is that,
instead of only considering the fusion of sets, it is also necessary
to track if it is necessary to break a set up into two smaller sets.
Putting everything from the previous sections together, we get an
outline of the algorithm for the general FLSA.

\begin{algorithm}[tbp]
\caption{General FLSA path algorithm}
\SetLine
\SetKwBlock{Initialize}{initialize}{end}
\Initialize{$\lambda_{2}=0$\;
$\beta_{k}=y_{k}$ for $k=1,\ldots,n$\;
$\tau_{kl}=0$ for $k,l=1,\ldots,n$; $k\neq l$\;
$F_{i}=\{i\}$ for $i=1,\ldots,n$\;
$n_{F}=n$\;
}%
\While{$n_{F}>1$}{Update $\beta_{F_{i}}$ and $\tau_{kl}$\;
Calculate the derivatives of $\beta_{F_{i}}$ w.r.t. $\lambda_{2}$
for $i=1,\ldots,n_{F}$\;
Solve the maximum flow problem for $F_{i}$ for $i=1,\ldots,n_{F}$\;
\eIf{not all flows from source are at capacity for graph $\tilde{\mathcal{G}}_{i}$}{Split
set $F_{i}$ into two smaller sets\;$n_{F}:=n_{F}+1$\;
}
{
Calculate next hitting time $h(\lambda_{2})$\;
Calculate the next violation time $v(\lambda_{2})$ when a set has
to be checked for violation of the constraints\;
\If{$h(\lambda_{2})<v(\lambda_{2})$}{Fuse the two sets that hit
each other\;$n_{F}:=n_{F}-1$\;}
}
Set $\lambda_{2}:=\min\left\{ h(\lambda_{2}),v(\lambda_{2})\right\} $\;
}
\label{GeneralFLSAAlg}
\end{algorithm}

It should be noted that this is a basic outline of the algorithm and
there is room for considerable efficiency gains when implementing
it. Most importantly, similar to the one-dimensional case from above,
the hitting times $h_{i,j}$ only have to be updated if either the
set $F_{i}$ or $F_{j}$ have changed since the last calculation.
The same is true for the maximum flow problems. The flows only have
to be updated if the underlying set has changed or a violation of
a constraint was triggered. Therefore, in every iteration only a small
number of sets is involved in the calculation and the computations
can be done quickly. Especially for larger sets, the computationally
most expensive step is solving the maximum flow problem. This gives
us the possibility to derive an approximate version of the algorithm
that is much faster as we will see in the simulations section below.

\subsection{Approximate algorithm}

The algorithm described above gives an exact solution to the problem.
However, for large sets of fused variables, the calculation of a maximal
flow is a bottleneck. In addition to this, if large sets of fused
variables split, the resulting sets tend to be very unequal in size,
often only splitting off a couple of nodes on the edges. Therefore,
a lot of time is spent on cases that do not influence the solution
very much. In order to speed up the algorithm, we propose
to not check sets of fused variables for splitting up once they are
larger than a certain size $K$. Then, for any set of size $K$ or
larger, the maximum flow problem does not have to be solved, saving
time on these especially computationally expensive sets. Also, as
the values of $\tau_{kl}$ are only used to determine when a set has
to be split, these also do not have to be updated any more for the
large sets. As we will see in the simulations section, the tradeoff
in accuracy for moderate values of $K$ is not very large but the
algorithm speeds up considerably.

\subsection{Simulations}

In order to evaluate the performance of our exact and approximate
algorithms above, we want to compare its speed and accuracy to the
approximate FLSA algorithm for the 2-dimensional case presented in
\citet{Friedman2007} as well as \emph{CVX} (see \citet{Grant2008a,Grant2008b}).
As we also want to compare the accuracy of the approximate algorithms,
we will use these techniques on simulated datasets, which we describe
in more detail below. 

The comparisons between the algorithms will be performed on
datasets of various sizes ranging from $10\times10$ to $200\times200$.
On each of these datasets, the solution will be computed for $50$
equally spaced values of $\lambda_{2}$ between $0$ and $0.5$. For
this speed comparison, there are two things to note. 

First, as noted before, \emph{CVX} cannot use the solution of a similar
$\lambda_{2}$ as a {}``warm start'' to speed up computation. So,
computing the solution for all $50$ values of $\lambda_{2}$ takes
roughly $50$ times as long as computing the solution for just one
value of $\lambda_{2}$. However, we chose to use it nonetheless as
it is an easy to use general convex solver that can handle sparse
matrices and is therefore equipped to handle large datasets. 

Second, the algorithm that is presented in this paper not only calculates
the solution at the $50$ values of $\lambda_{2}$, but at all breakpoints
of the piecewise-linear solution. The whole path is saved in a compact
format and can be used to extract the solution at other values of
$\lambda_{2}$ later much faster. For our dataset here, the solution
path has at least as many breakpoints as there are datapoints, i.e.
$n^{2}$ for an $n\times n$ grid. However, we still only let the
other algorithm evaluate it for $50$ values of $\lambda_{2}$ as
we deemed it unrealistic that the solution for possibly thousands
of $\lambda_{2}$ values is needed.

\subsubsection{The dataset}

\begin{figure}
\begin{centering}
\includegraphics[width=10cm]{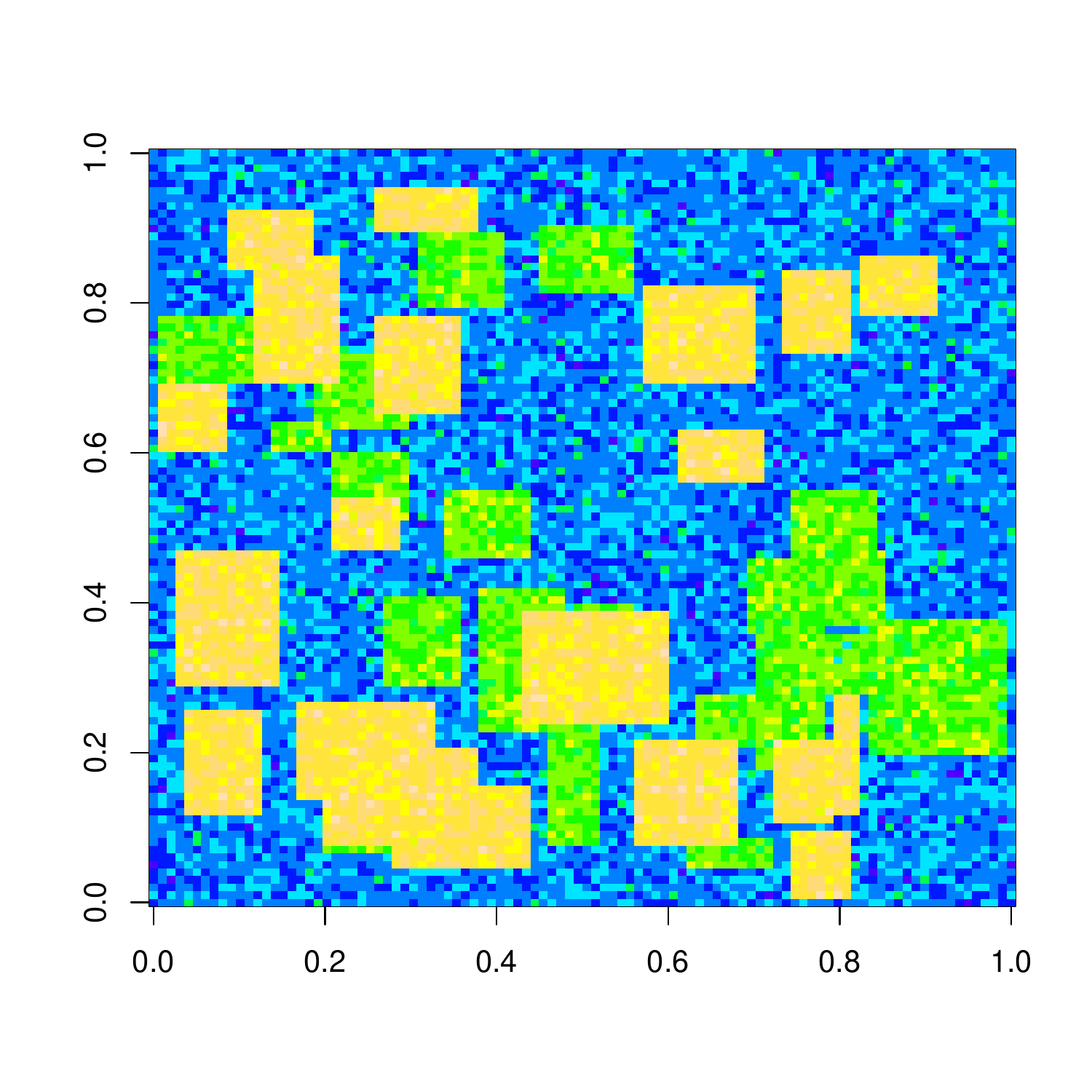}
\caption[Example image of 2-dim dataset]{A sample image of the simulated 2-dimensional dataset\label{fig:Simulated2Dim}}
\end{centering}
\end{figure}

In our comparisons, we want to use the 2-dimensional FLSA. Therefore,
our data will consist of data $\by=\left\{ y_{kl}\right\} $ with $k=1,\ldots,n_{1}$
and $l=1,\ldots,n_{2}$ with corresponding coefficients $\beta_{kl}$.
The difference of coefficients will be penalized if they are neighbors
on the 2-dimensional grid (horizontal or vertical),
i.e. the loss function we want to minimize is
\[
L(\by,\bbeta)=\frac{1}{2}\sum_{k=1}^{n_{1}}\sum_{l=1}^{n_{2}}(y_{kl}-\beta_{kl})^{2}+\sum_{k=1}^{n_1-1}\sum_{l=1}^{n_{2}}|\beta_{kl}-\beta_{k+1,l}|+\sum_{k=1}^{n_{1}}\sum_{l=1}^{n_2-1}|\beta_{kl}-\beta_{k,l+1}|.
\]
where we have already set $\lambda_1=0$.
As shown above, we can get the solution for any $\lambda_{1}$ by
soft-thresholding the solution for $\lambda_{1}=0$. In our simulated
dataset, we set $n_{1}=n_{2}=n$ for various values of $n$. The value
of $y_{kl}$ is being generated as follows:

\begin{enumerate}
\item Set $y_{kl}=0$ for all $k,l=1,\ldots,n$.
\item For some rectangles of random size, change the value of $y$ to either
1 or 2, such that roughly 20\% have value 1 and 20\% have value 2.
\item To every point $y_{kl}$ add standard normal noise with standard deviation
of 0.2.
\end{enumerate}
A sample image of what the simulated dataset looks like can be seen
in Figure \ref{fig:Simulated2Dim}.

\subsubsection{Results}

First, we compare the computation time of the three methods. The results
can be seen in Table \ref{tab:2DimFLSACompTime}. The path algorithm
as well as the component-wise algorithm are both much faster than
the general convex solver CVX. When comparing the path algorithm to
the component-wise algorithm, we see that they have roughly the same
speed for low values of $K$, except for the small $10\times10$ dataset. 

With respect to accuracy, we measure both the sup-norm error as well
as the root mean squared deviation (RMSD). These measures are being
calculated for each value of $\lambda_{2}$ and the largest values,
averaged over several simulation runs are being displayed in Tables \ref{tab:2DimFLSAAbsError}
and \ref{tab:2DimFLSARMSD}. CVX returns the exact solution in all
cases, although at the cost of a rather slow speed. The component-wise
algorithm on the other hand is quite fast, however only yields an
approximate solution, although with a rather small error rate in
terms of RMSD. With varying $K$, the path algorithm is in between
the other two methods. However, for values of $K$ around $500-1000$,
the path algorithm is very accurate but still a lot faster than CVX.
For most practical application, this time-accuracy tradeoff may be
worthwhile. 

\begin{table}
\begin{centering}
\begin{tabular}{|c|c|c|c|c|c|}
\hline 
\multicolumn{2}{|c|}{Image size} & $10\times10$ & $50\times50$ & $100\times100$ & $200\times200$\tabularnewline
\hline
\hline 
\multicolumn{2}{|c|}{CVX} & 36 & 140 & 1000 & 6800\tabularnewline
\hline 
\multicolumn{2}{|c|}{Component-wise Alg.} & 0.062 & 0.61 & 2.2 & 7.7\tabularnewline
\hline 
\multirow{11}{*}{Path Alg.} & $K=1$ & 0.0031 & 0.17 & 1.1 & 10\tabularnewline
 & $K=2$ & 0.0032 & 0.18 & 1.1 & 10\tabularnewline
 & $K=5$ & 0.0040 & 0.20 & 1.3 & 10\tabularnewline
 & $K=10$ & 0.0046 & 0.23 & 1.4 & 11\tabularnewline
 & $K=50$ & 0.0077 & 0.37 & 2.0 & 14\tabularnewline
 & $K=100$ & - & 0.51 & 3.0 & 18\tabularnewline
 & $K=500$ & - & 1.3 & 11 & 62\tabularnewline
 & $K=1000$ & - & 1.4 & 21 & 120\tabularnewline
 & $K=2000$ & - & 1.4 & 22 & 290\tabularnewline
 & $K=5000$ & - & - & 23 & 400\tabularnewline
 & exact & 0.0079 & 1.4 & 24 & -\tabularnewline
\hline
\end{tabular}
\caption[Speed comparison for 2-dimensional FLSA]{Speed comparison for the 2-dimensional FLSA in seconds. The solution is evaluated for 50 values of $\lambda_{2}$ between 0 and 0.5. The results are averaged over 10 runs for the $10\times 10$ and $50\times 50$ dataset and 4 simulation runs for the rest.}
\label{tab:2DimFLSACompTime}
\end{centering}
\end{table}

\begin{table}
\begin{centering}
\begin{tabular}{|c|c|c|c|c|c|}
\hline 
\multicolumn{2}{|c|}{Image size} & $10\times10$ & $50\times50$ & $100\times100$ & $200\times200$\tabularnewline
\hline
\hline 
\multicolumn{2}{|c|}{CVX} & 0 & 0 & 0 & 0\tabularnewline
\hline 
\multicolumn{2}{|c|}{Component-wise Alg.} & 0.30 & 0.80 & 0.73 & 1.0\tabularnewline
\hline 
\multirow{11}{*}{Path Alg.} & $K=1$ & 0.31 & 0.81  & 0.73 & 1.0\tabularnewline
 & $K=2$ & 0.26 & 0.75 & 0.73 & 1.0\tabularnewline
 & $K=5$ & 0.22 & 0.74 & 0.73 & 0.90\tabularnewline
 & $K=10$ & 0.15 & 0.68 & 0.73 & 0.90\tabularnewline
 & $K=50$ & 0 & 0.44 & 0.22 & 0.52\tabularnewline
 & $K=100$ & 0 & 0.35 & 0.18 & 0.43\tabularnewline
 & $K=500$ & 0 & 0.025 & 0.047 & 0.14\tabularnewline
 & $K=1000$ & 0 & 0 & 0.013 & 0.12\tabularnewline
 & $K=2000$ & 0 & 0 & 0.00017 & 0.050\tabularnewline
 & $K=5000$ & 0 & 0 & 0.00017 & 0.021\tabularnewline
 & exact & 0 & 0 & 0 & -\tabularnewline
\hline
\end{tabular}
\caption[Absolute error for 2-dimensional FLSA]{Absolute error accuracy comparison for the 2-dimensional FLSA.
The accuracy of the approximate version of the path algorithm and
the component-wise algorithm are compared to the exact solution using
the supremum norm. The largest error of the 50 values of $\lambda_{2}$
is reported. The results are averaged over 10 runs for the $10\times 10$ and $50\times 50$ dataset and 4 simulation runs for the rest.}
\label{tab:2DimFLSAAbsError}
\end{centering}
\end{table}

\begin{table}
\begin{centering}
\begin{tabular}{|c|c|c|c|c|c|}
\hline 
\multicolumn{2}{|c|}{Image size} & $10\times10$ & $50\times50$ & $100\times100$ & $200\times200$\tabularnewline
\hline
\hline 
\multicolumn{2}{|c|}{CVX} & 0 & 0 & 0 & 0\tabularnewline
\hline 
\multicolumn{2}{|c|}{Component-wise Alg.} & 0.056 & 0.066 & 0.041 & 0.030\tabularnewline
\hline 
\multirow{11}{*}{Path Alg.} & $K=1$ & 0.059 & 0.067 & 0.045 & 0.031\tabularnewline
 & $K=2$ & 0.056 & 0.066 & 0.044 & 0.030\tabularnewline
 & $K=5$ & 0.041 & 0.059 & 0.042 & 0.029\tabularnewline
 & $K=10$ & 0.029 & 0.053 & 0.039 & 0.027\tabularnewline
 & $K=50$ & $<10^{-5}$ & 0.035 & 0.027 & 0.020\tabularnewline
 & $K=100$ & 0 & 0.022 & 0.022 & 0.015\tabularnewline
 & $K=500$ & 0 & 0.0050 & 0.0059 & 0.0078\tabularnewline
 & $K=1000$ & 0 & $<10^{-5}$ & 0.00026 & 0.0053\tabularnewline
 & $K=2000$ & 0 & $<10^{-5}$ & $<10^{-5}$ & 0.0028\tabularnewline
 & $K=5000$ & 0 & 0 & $<10^{-5}$ & 0.0019\tabularnewline
 & exact & 0 & 0 & 0 & -\tabularnewline
\hline
\end{tabular}
\caption[RMSD for 2-dimensional FLSA]{Root mean square deviation (RMSD) accuracy comparison for the 2-dimensional
FLSA. The accuracy of the approximate version of the path algorithm
and the component-wise algorithm are compared to the exact solution
using the RMSD. The largest error of the 50 values of $\lambda_{2}$
is reported. The results are averaged over 10 runs for the $10\times 10$ and $50\times 50$ dataset and 4 simulation runs for the rest.}
\label{tab:2DimFLSARMSD}
\end{centering}
\end{table}

\section{Conclusion\label{sec:Conclusion}}

In this article we develop a path algorithm for the Fused Lasso Signal Approximator
in its one-dimensional and general form. We compared the speed and accuracy of the FLSA algorithm to other
available methods and conclude that our method has advantages in terms
of speed and the amount of information gathered and stored. Especially
compared to standard convex solvers, our path algorithm is much faster
for the FLSA. It is also very easy and quick to extract results for additional
penalty parameter values. 
We also extend this work to the case of the general Fused Lasso where we restrict
ourselves to predictor matrices with $\text{rank}(X)=p$ where $X \in \mathds{R}^{n\times p}$
(see the Online Supplement).

Apart from the work presented here, there are several ways how we
plan to expand on it in the future. It is possible to expand
the Fused Lasso by allowing each summand in the penalty terms to have
separate weights, i.e. a loss function of the form 

\[
L_{\lambda_{1},\lambda_{2}}(y,X,\beta)=\frac{1}{2}(y-X\beta)^{T}(y-X\beta)+\lambda_{1}\sum_{k=1}^{p}w_{k}|\beta_{k}|+\lambda_{2}\sum_{(k,l)\in E,k<l}w_{kl}|\beta_{k}-\beta_{l}|.\]
This more complicated model can be solved by a generalization of the
algorithms presented in this article. Also, the current algorithm for the FLSA is not optimized form
graphs with large number of edges in them. We plan to develop a version that takes cliques into account to
achieve further efficiency gains.

We hope that our algorithms will be used to analyze data and as a
building block for other new models. In order to facilitate this we will be publishing implementations
of the algorithms in the form of an R package on CRAN and the authors website.

\bibliographystyle{plainnat}
\bibliography{hhoeflin}

\end{document}


\title{Supplement to ``A path algorithm for the Fused Lasso Signal Approximator''}
\author{Holger Hoefling\thanks{Email: hhoeflin@gmail.com} \\ Stanford University}

\maketitle

\section{Introduction}
In this online supplement to the article ``A path algorithm for the Fused Lasso'', we will give proofs and
further results that would have gone beyond the possible size and scope of the paper. In Section \ref{sec:OneDimFLSA}
we will discuss the computational and memory complexity of the algorithm. In addition to this we will give an alternative proof to Theorem 2 of the article in which we show that for increasing $\lambda_2$, fused sets of variables do not become unfused again. In Section \ref{sec:GeneralFLSA} we will give the proofs to Theorem 3 and Proposition 1 that complete our results for the general Fused Lasso Signal Approximator. Finally, in Section \ref{sec:GeneralFusedLasso}, we will extend the path algorithm of the FLSA to the case of the general Fused Lasso where for the matrix of predictors $\bX \in \mathds{R}^{n\times p}$ we have $\text{rank}(\bX)=p$ and we also present the associated proofs.

\section{One dimensional FLSA \label{sec:OneDimFLSA}}
\subsection{Computational complexity}
In Section 2 of the article, we have presented the algorithm that solves the one-dimensional Fused Lasso Signal Approximator. We stated that the computational complexity of the algorithm is $n\log(n)$. In order to calculate the complexity of the algorithm,
first note that the initialization step takes $2n+2$ operations.
However, most of the computations are performed for calculating the
next hitting time $h(\lambda_{2})$. When $\lambda_{2}=0$, the derivative
$\partial\beta_{F_{i}}/\partial\lambda_{2}$ has to be computed
for all $n$ groups and the smallest hitting time identified. This
takes on the order $O(n)$ operations. However, in subsequent steps,
$h_{i,i+1}$ remain the same except if either $F_{i}$ or $F_{i+1}$
was fused with its neighbor in the last iterations. This way, in each
iteration only 2 values of $h_{i,i+1}$ have to be updated. Finding
the smallest value of $h_{i,i+1}$ requires operations on the order
$O(\log(n))$ if an efficient data structure is used to save the $h_{i,i+1}$
(e.g. a binary tree). The same is true for the updates of the $\beta_{F_{i}}$.
Only the $\beta_{F_{i}}$ of the groups that have just been fused
have to be updated. For the other groups $\partial\beta_{F_{i}}/\partial\lambda_{2}$
stays the same, so we can interpolate $\beta_{F_{i}}(\lambda_{2})$
when we need it based on $\lambda_{2}^{0}$ when this group was created
as well as $\beta_{F_{i}}(\lambda_{2}^{0})$ and its derivative. So
this also only involves a constant number of operations per iteration.
As with every iteration, the number of sets $n_{F}$ decreases by
one, there are at most $n-1$ iterations. Therefore, the entire solution
path can be obtained with a computational complexity on the order
$O(n\log(n))$.

\subsection{Storing the solution path efficiently}

One possibility to store and retrieve the optimal coefficients for
a value of $\lambda_{2}$ would be to store the whole coefficient vector $\bbeta$ for every value of $\lambda_{2}$ for which 2 sets
are fused and interpolate linearly in between. However, then it would
be necessary to store $n^{2}$ entries, which would be impossible
for large values of $n$. However, when done efficiently, it is possible
to store the entire solution with a memory requirement that only grows
linearly in $n$. For this, the result is stored in the form of a
binary tree. Every node contains a value for $\lambda_{2}$ at which
this node (corresponding to a set of fused variables) becomes active
and the correct value of $\beta_{F_{i}}$ for this same value $\lambda_{2}$. 

\begin{figure}[t]
\begin{centering}
\includegraphics[viewport=0bp 150bp 780bp 540bp,clip,width=12cm]{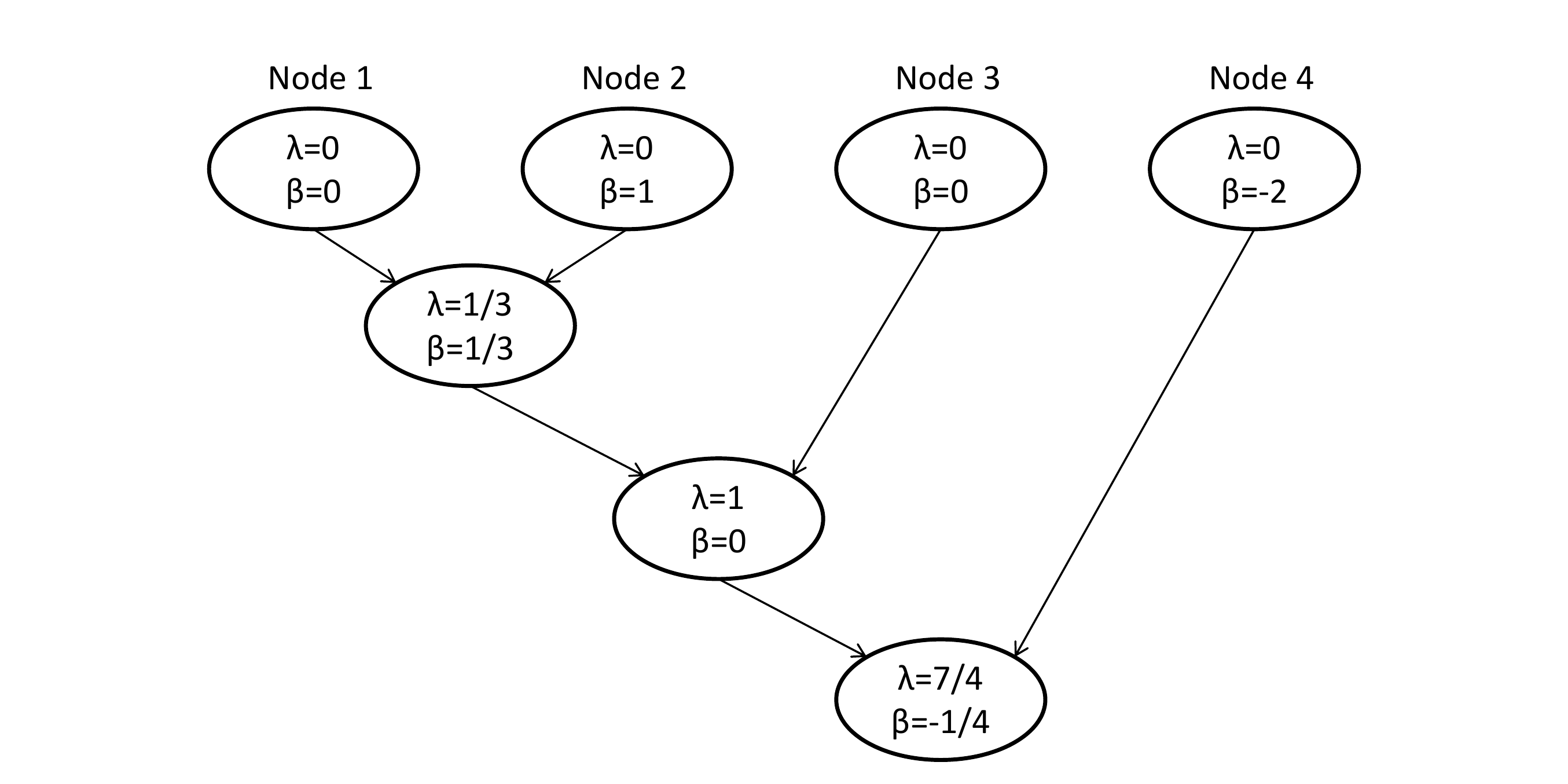}
\caption[Example tree for 1-dim FLSA]{Example of a tree for storing the solution path of the one-dimensional
FLSA.}
\label{fig:ExampleTreeFLSA1Dim}
\end{centering}
\end{figure}

We start by creating $n$ unconnected nodes corresponding to the $n$
coefficients for $\lambda_{2}=0$ and $\beta_{k}=y_{k}$. Then when
two sets of coefficients fuse, add a new node that is the parent of
the two nodes corresponding to the 2 groups that were just fused.
In the new node, store the value of $\lambda_{2}$ as well as the
updated value of the coefficient $\beta_{F_{i}}$. At the end of the
algorithm, we will have one binary tree (for a small example see Figure
\ref{fig:ExampleTreeFLSA1Dim}). In order to retrieve one value of
the solution at $\tilde{\lambda}_{2}$ for coefficient $\beta_{k}$,
first start at the leaf node of coefficient $\beta_{k}$. Climb up
the tree until $\lambda_{2}$ of the next parent node is larger than
$\tilde{\lambda}_{2}$. The correct solution is then a linear interpolation
between the $\beta_{F_{i}}$ values stored in the current node and
the parent node. Using this method, the complexity for looking the
whole vector of solutions $\beta$ for a particular value of $\lambda_{2}$
is then also $O(n\log n)$. Therefore, running the path algorithm
and getting a solution vector has a total complexity of $O(n\log n)$.
We will also be able to see this in the next section when we compare
the speed of this algorithm to other existing methods for this problem. 

\subsection{Alternative proof of Theorem 2}\label{AppOneDimFLSA}
An important prerequisite for the algorithm is the result of Theorem 2 in the article. A proof has already been given in  \citet{Friedman2007} and here we want to give an alternative version. It is based on the results of the Genreal Fused Lasso Signal Approximator, so read Section 3 of the article before this proof.

\begin{thm}{2}
For the one-dimensional FLSA assume that for some $\lambda_{2}^{0}>0$
and $\lambda_{1}=0$, we have that $\beta_{k}(\lambda_{2}^{0})=\beta_{k+1}(\lambda_{2}^{0})$
for some $1\le k\le N-1$. Then, for any $\lambda_{2}>\lambda_{2}^{0}$
we have that $\beta_{k}(\lambda_{2})=\beta_{k+1}(\lambda_{2})$.
\end{thm}

\begin{proof}First note that in Section 3.2.2,
we described the maximum flow problem that we have to solve and under
which conditions it is necessary to break up the set of fused variables.
In the case of the one-dimensional FLSA, the maximum flow graph is
especially simple. Assume that an arbitrary set of fused variables
$|F|$ is $\{ k_{0},k_{0}+1,\ldots,\allowbreak k_{0}+|F|-1\} $, which
has the edges $\{ (k_{0},k_{0}+1),\allowbreak \ldots ,\allowbreak (k_{0}+|F|-2,k_{0}+|F|-1)\}$.
For the capacities on these edges, it is sufficient to note that they
are always $\ge1$. Now it only remains to specify which edges are
connected to the source node $r$ and the sink node $s$ and which
capacities they have. Essentially, there are 4 possible situations:

\begin{description}
\item [Case~1 $\beta_{k_{0}-1},\beta_{k_{0}+|F|}>\beta_{F}$:] In this
case we have that $\frac{\partial\beta_{F}}{\partial\lambda_{2}}=\frac{2}{|F|}$.
Then we have that $k_{0}$ and $k_{0}+|F|-1$ are connected to the
source $r$ and all other nodes are connected to the sink $s$. For
the capacities from the source, we have $c_{rl}=1-\frac{2}{|F|}\le1$
and for the capacities to the sink it holds that $c_{ks}=\frac{2}{|F|}\le1$.
As the capacities on the inner nodes is at least 1 as stated above,
it is easy to see that for the maximum flow we always have $f_{rl}=c_{rl}$
for $l=k_{0},k_{0}+|F|-1$ and therefore the group will never be split.
\item [Case~2 $\beta_{k_{0}+1}>\beta_{F}>\beta_{k_{0}+|F|}$:] Here,
we have that $\frac{\partial\beta_{F}}{\partial\lambda_{2}}=0$ and
node $k_{0}$ is connected to the source $r$ whereas $k_{0}+|F|-1$
is connected to the sink $s$. All other nodes are not connected to
either the source or the sink. For the capacities in this case we
have $c_{rk_{0}}=c_{k_{0}+|F|-1,s}=1$ and therefore the maximum flow
is always $f_{rk_{0}}=f_{k_{0},k_{0}+1}=\cdots=f_{k_{0}+|F|-2,k_{0}+|F|-1}=f_{k_{0}+|F|-1,s}=1$
and therefore, the group will never be split. 
\item [Case~3 $\beta_{k_{0}+1}<\beta_{F}<\beta_{k_{0}+|F|}$:] Similar
to case 2.
\item [Case~4 $\beta_{k_{0}-1},\beta_{k_{0}+|F|}<\beta_{F}$:] Similar
to case 1.
\end{description}
In these cases we assumed that $k_{0}\neq1$ and $k_{0}+|F|-1\neq N$,
i.e. that the set of fused variables is not at the boundary of the
graph. However, it is easy to see that in this case the set will also
not break apart for increasing $\lambda_{2}$. Therefore, the proposition
holds.\end{proof}

\section{The general Fused Lasso Signal Approximator \label{sec:GeneralFLSA}}
In this section we want to give the proofs of Theorem 3 and Proposition 1 of the article. First, we will give
the proof of Theorem 3, then the one of Proposition 1.
\subsection{Proof of Theorem 3}
\label{AppProofSolutionLinear}
\begin{proof}In order to show that this is indeed the solution, we
have to show that for any $\lambda_{2}\in[\lambda_{2}^{0},\lambda_{2}^{0}+\Delta]$,
the subgradient Equations (5) hold. We will do
this for every group $F_{i}$ separately. We know that \[
\frac{\partial\beta_{F_{i}}}{\partial\lambda_{2}}=-\frac{1}{|F_{i}|}\sum_{j\neq i}\sum_{(k,l)\in E;k\in F_{i};l\in F_{j}}t_{kl}.\]

For the maximum flow graph, we know that all nodes except the source
and the sink have a net flow of 0. Furthermore, we also know that
the sum of all capacities coming from the source is the same as the
sum of all capacities going to the sink as\[
\sum_{(r,l)\in\tilde{E}_{i}}c_{rl}-\sum_{(k,s)\in\tilde{E}_{i}}c_{ks}=\sum_{k\in F_{i}}p_{k}=0\]
where we use that for any node that is not connected to either the
source or the sink, $p_{k}=0$. From this we can see that if all flows
coming from the source are at maximum capacity, so are all flows going
to the sink. Therefore, for a maximum flow with all flows from the
source being at maximum capacity, we have\[
\sum_{l:(k,l)\in E;l\in F_{i}}f_{kl}=\begin{cases}
-f_{kr}=c_{rk}=p_{k} & \mbox{if }(r,k)\in\tilde{E}_{i}\\
-f_{ks}=-c_{ks}=p_{k} & \mbox{if }(k,s)\in\tilde{E}_{i}\\
0=p_{k} & \mbox{otherwise}\end{cases}\]
so that overall we get\begin{align*}
\beta_{k}(\lambda_{2})-y_{k}+\lambda_{2}\sum_{l:(k,l)\in E}t_{kl}(\lambda_{2}) & \displaybreak[0] \\
=\beta_{k}(\lambda_{2}^{0})-y_{k}+\lambda_{2}\sum_{l:(k,l)\in E}t_{kl}(\lambda_{2}^{0})+\frac{\partial\beta_{F_{i}}}{\partial\lambda_{2}}(\lambda_{2}-\lambda_{2}^{0}) & \displaybreak[0]\\
+\sum_{l\in F_{i}:(k,l)\in E}f_{kl}(\lambda_{2}-\lambda_{2}^{0})+\sum_{l\not\in F_{i}:(k,l)\in E}t_{kl}(\lambda_{2}^{0})(\lambda_{2}-\lambda_{2}^{0}) & \displaybreak[0]\\
=(\lambda_{2}-\lambda_{2}^{0})\left[\frac{\partial\beta_{F_{i}}}{\partial\lambda_{2}}+\sum_{l\in F_{i}:(k,l)\in E}f_{kl}+\sum_{l\not\in F_{i}:(k,l)\in E}t_{kl}(\lambda_{2}^{0})\right] & \displaybreak[0]\\
=(\lambda_{2}-\lambda_{2}^{0})\left[\sum_{l\in F_{i}:(k,l)\in E}f_{kl}-p_{k}\right] & =0\end{align*}
where we use the definition of $p_{k}(\lambda_{2}^{0})=-\sum_{l\not\in F_{i}:(k,l)\in E}t_{kl}(\lambda_{2}^{0})-\frac{\partial\beta_{F_{i}}}{\partial\lambda_{2}}$
in the second to last equality. As this is true for every $k\in F_{i}$
and group $F_{i}$, the solution as proposed in the theorem solves
the subgradient equations and therefore minimizes the loss function. 
\end{proof}

\subsection{Proof of Proposition 1}
\label{AppProofFuseSplitSets}

\begin{proof}First, we want to show that after the fusion and splitting
steps finish, we have a valid set of fused variables. For this, we
have to show that for sets $F_{i}$ and $F_{j}$ it holds that $\beta_{F_{i}}(\lambda_{2})\neq\beta_{F_{j}}(\lambda_{2})$
for $\lambda_{2}\in(\lambda_{2}^{0},\lambda_{2}^{0}+\varepsilon)$
for some $\varepsilon>0$. However, this is certainly true. If $\beta_{F_{i}}(\lambda_{2}^{0})\neq\beta_{F_{j}}(\lambda_{2}^{0})$,
then this holds also for some interval $\lambda_{2}\in(\lambda_{2}^{0},\lambda_{2}^{0}+\varepsilon)$
as $\beta_{k}(\lambda_{2})$ is continuous in $\lambda_{2}$. If we
have that $\beta_{F_{i}}(\lambda_{2}^{0})=\beta_{F_{j}}(\lambda_{2}^{0})$,
then we can assume that $\frac{\partial\beta_{F_{i}}}{\partial\lambda_{2}}(\lambda_{2}^{0})>\frac{\partial\beta_{F_{j}}}{\partial\lambda_{2}}(\lambda_{2}^{0})$
and $t_{kl}=1$ for $k\in F_{i},l\in F_{j};(k,l)\in E$, as otherwise
$F_{i}$ and $F_{j}$ would have been fused. However, this means that
$\beta_{F_{i}}(\lambda_{2})>\beta_{F_{j}}(\lambda_{2})$ for $\lambda_{2}\in(\lambda_{2}^{0},\lambda_{2}^{0}+\varepsilon)$
and that $t_{kl}(\lambda_{2})=\mbox{\mbox{sign}}(\beta_{F_{i}}(\lambda_{2})-\beta_{F_{j}}(\lambda_{2}))$,
so the grouping is valid.

Second, for all of the sets at the end of the splitting steps, the
maximal flow condition of Theorem 3 trivially
holds as any set for which it doesn't hold, would be split up into
two smaller sets (and it always holds for sets of size 1). 

Therefore, it only remains to show that the fusion and splitting steps
converge after a finite number of iterations. For this, we show that
if a set $F_{i}$ was split into subgroups $R_{i}$ and $S_{i}$ at
penalty parameter $\lambda_{2}^{0}$, then $R_{i}$ and $S_{i}$ will
not be merged at $\lambda_{2}^{0}$ in a subsequent iteration. From
this, we will conclude that there is only a finite number of possible
iterations, as there is only a finite number of possible sets and
we cannot have infinite cycles of fusions and splits for the sets.
Therefore, the algorithm converges after a finite number of steps.

So, as $F_{i}$ was split into $R_{i}$ and $S_{i}$, we know that
$R_{i}\neq\emptyset$ as well as $S_{i}\neq\emptyset$. For set $R_{i}$
consider the capacities of edges the source into $R_{i}$ minus the
capacities of edges going from $R_{i}$ either to $S_{i}$ or directly
to the sink. As we split group $F_{i}$, we know that the capacities
going into $R_{i}$ are larger than the capacities going out and therefore,\[
\sum_{(r,l)\in\tilde{E}_{i};l\in R_{i}}c_{rl}>\sum_{(k,l)\in\tilde{E}_{i};k\in R_{i},l\in S_{i}}c_{kl}+\sum_{(k,s)\in\tilde{E}_{i};k\in R_{i}}c_{ks}.\]

For all edges $(k,l)\in\tilde{E}_{i}$ with $k\in R_{i}$ and $l\in S_{i}$
we also know that $1=c_{kl}=f_{kl}=t_{kl}$. Here $f_{kl}=c_{kl}$
follows as otherwise $l$ would be connected to the source in the
residual graph and therefore $l\in R_{i}$, which is not the case
as $l\in S_{i}$. Also, it follows that $c_{kl}=1$, as $c_{kl}=\infty$
is the only other option which cannot hold as the maximum flow is
finite. Furthermore $t_{kl}=1$ as $c_{kl}=1$. Then\[
\sum_{(r,l)\in\tilde{E}_{i};l\in R_{i}}c_{rl}-\sum_{(k,s)\in\tilde{E}_{i};k\in R_{i}}c_{ks}-\sum_{(k,l)\in\tilde{E}_{i};k\in R_{i},l\in S_{i}}t_{kl}>0.\]

Using that $c_{rl}=p_{l}$ for $(r,l)\in\tilde{E}_{i}$ and $c_{ks}=-p_{k}$
for $(k,s)\in\tilde{E}_{i}$, we get
\begin{equation}
\sum_{k\in R_{i}}p_{k}-\sum_{(k,l)\in\tilde{E}_{i};k\in R_{i},l\in S_{i}}t_{kl}>0.\label{eq:ProofFuseSplitEq1}\end{equation}
Let $p_{k}^{R_{i}}$ be the push on node $k$ in the graph associated
with group $R_{i}$ and $p_{l}^{S_{i}}$ the push on node $l$ associated
with $S_{i}$ after the split. From equation (8),
we know that \[
\sum_{k\in R_{i}}p_{k}^{R_{i}}=0.\]

We can also infer from the definition of the push on $F_{i}$ and
$R_{i}$ that 
\[
\sum_{k\in R_{i}}p_{k}^{R_{i}}+|R_{i}|\frac{\partial\beta_{R_{i}}}{\partial\lambda_{2}}=\sum_{k\in R_{i}}p_{k}-\sum_{k\in R_{i},l\in S_{i,}(k,l)\in E}t_{kl}+|R_{i}|\frac{\partial\beta_{F_{i}}}{\partial\lambda_{2}}\]
and therefore using equation (\ref{eq:ProofFuseSplitEq1}) and $\sum_{k\in R_{i}}p_{k}^{R_{i}}=0$
that
\[
\frac{\partial\beta_{R_{i}}}{\partial\lambda_{2}}>\frac{\partial\beta_{F_{i}}}{\partial\lambda_{2}}.\]
Now, again using equation (8), we get
\[
\frac{\partial\beta_{F_{i}}}{\partial\lambda_{2}}=\frac{|R_{i}|}{|R_{i}|+|S_{i}|}\frac{\partial\beta_{R_{i}}}{\partial\lambda_{2}}+\frac{|S_{i}|}{|R_{i}|+|S_{i}|}\frac{\partial\beta_{S_{i}}}{\partial\lambda_{2}}\]
and thus
\[
\frac{\partial\beta_{R_{i}}}{\partial\lambda_{2}}>\frac{\partial\beta_{F_{i}}}{\partial\lambda_{2}}>\frac{\partial\beta_{S_{i}}}{\partial\lambda_{2}}.\]
As also $t_{kl}=1$ for $k\in R_{i}$ and $l\in S_{i}$, we can see
that the groups that have just been split at $\lambda_{2}^{0}$ cannot
be merged again at the same penalty parameter $\lambda_{2}^{0}$.

In turn, this also means that any two sets that have just been fused,
cannot be immediately split up again into the same original sets at
$\lambda_{2}^{0}$ as this would lead us back to the starting position,
which would force us to fuse the two sets again. However, we have
just shown that this cannot happen and therefore we also 
see that any two sets that are being fused, are not being split up
again immediately.
\end{proof}

\section{Path algorithm for the General Fused Lasso\label{sec:GeneralFusedLasso}}
In this section we want to expand the result of the article from a path algorithm of the
general FLSA to a path algorithm for the general Fused Lasso with a predictor matrix $\bX\in \mathds{R}^{n \times p}$
where $\text{rank}(\bX)=p$.
The loss function we want to minimize in this case is
\[
L_{\lambda_{1},\lambda_{2}}(\by,\bX,\bbeta)=\frac{1}{2}(\by-\bX\bbeta)^{T}(\by-\bX\bbeta)+\lambda_{1}\sum_{k=1}^{p}|\beta_{k}|+\lambda_{2}\sum_{(k,l)\in E,k<l}|\beta_{k}-\beta_{l}|
\]
where $\bX\in \mathds{R}^{n\times p}$ and $\mathcal{G}=(V,E)$ a graph that
defines the penalty structure on the difference of coefficients $\beta_{k}-\beta_{l}$
for $(k,l)\in E$, the set of undirected edges. One important difference
to the $\bX=\bI$ case is that here, we cannot get the solution for all
values of $\lambda_{1}$ by soft-thresholding the solutions for $\lambda_{1}=0$
as we did for the FLSA. Therefore, we cannot calculate one path from
which it is easy to get all solutions for any combinations of $(\lambda_{1},\lambda_{2})$.
Instead, we will restrict ourselves to calculate the whole solution
path for $\lambda_{2}$ for a fixed value $\lambda_{1}$. Therefore,
we have to incorporate the additional penalty $\lambda_{1}\sum_{k=1}^{p}|\beta_{k}|$
into the algorithm. This will be done in a fashion similar to the
LARS algorithm (see \citet{Efron2002}) by having active and inactive
sets of fused variables. 

\begin{figure}
\begin{centering}
\includegraphics[width=8cm]{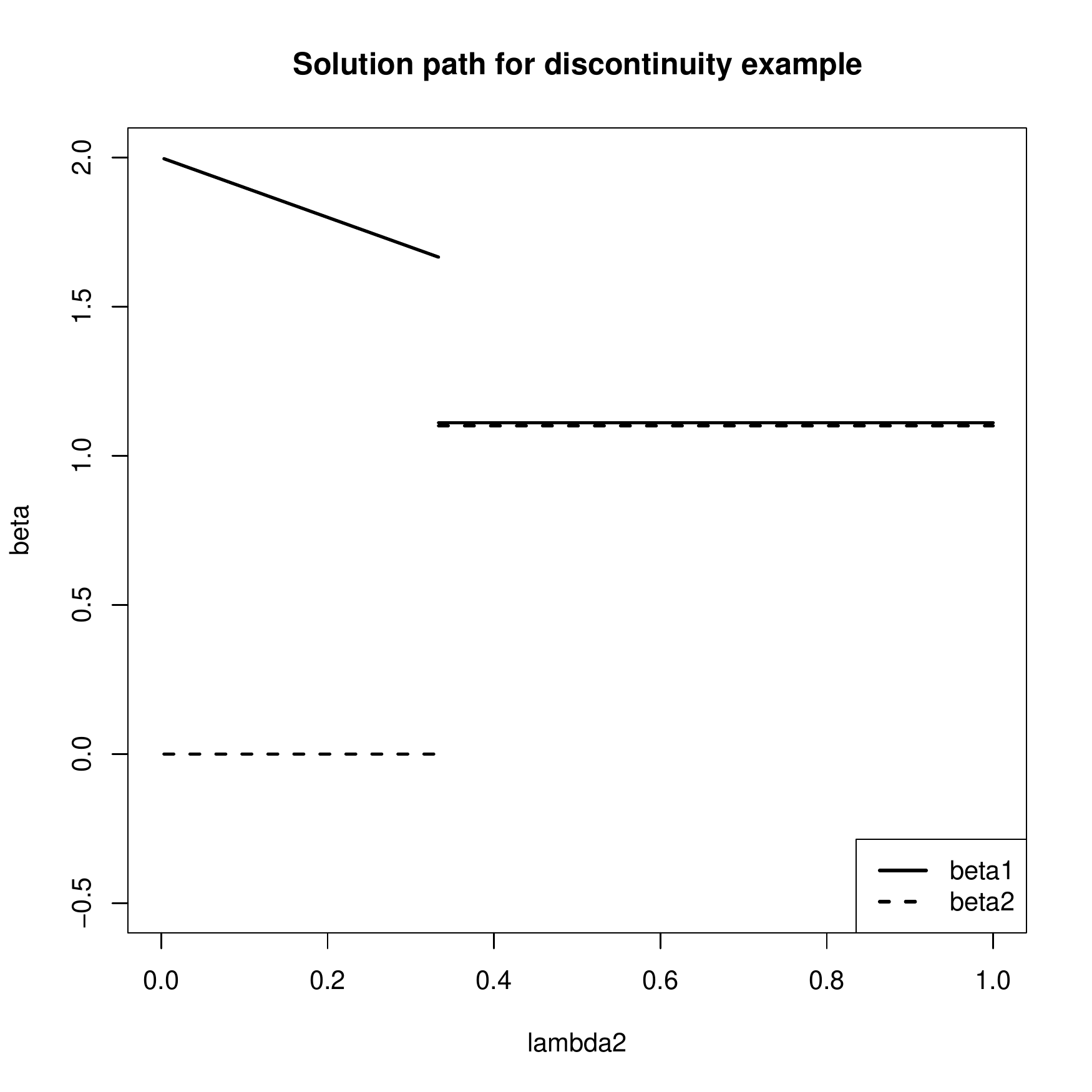}
\caption[Example of discontinuous solution path]{Example of a discontinuous solution path if $\text{rank}(X)<p$.}
\label{fig:exampleDiscontinuous}
\end{centering}
\end{figure}

The reason that we restrict the development of the algorithm to the case of
$\text{rank}(X)=p$ is that if $\text{rank}(X)<p$, then the resulting solution
path w.r.t. $\lambda_2$ may have discontinuities. Here is a very small example where
this happens. Assume that $n=1, p=2$ with $y_1=3$ as well as $x_{11}=1$ and $x_{12}=0.5$. 
For $\lambda_1=1$, the solution path for $\lambda_2$ can be seen in Figure \ref{fig:exampleDiscontinuous}.
As we can see, for $\lambda_2=1/3$, the solution path has a jump.

In this section, we will first enhance the definition of sets of fused
variables to incorporate that these sets can now also be active or
inactive. Next, we will show that the solution path is again piecewise
linear and how to calculate its slope and the necessary changes at
the breakpoints. We will end this section by giving the complete algorithm
that calculates the solution path for $\lambda_{2}$ where $\lambda_{1}$
is assumed to be fixed.

\subsection{Sets of fused variables and active sets}

In the case of the general Fused Lasso, it is not possible to assume
that $\lambda_{1}=0$ and therefore we have to account for the additional
penalty $\lambda_{1}\sum_{k=1}^{p}|\beta_{k}|$ in the loss function.
Therefore, in addition to sets of fused variables as in the case of
the FLSA algorithm, it is also necessary to keep track if these sets
$F_{i}$ are currently {}``active'', that is have an associated
coefficient $\beta_{F_{i}}\neq 0$, or inactive when $\beta_{F_{i}}=0$.
In order to account for this,
we have to adapt the definition of sets of variables. However, before
we do this we should have a look at the $k$-th subgradient equation
for the general loss function first:
\begin{equation}
\frac{\partial L_{\lambda_{1},\lambda_{2}}}{\partial\beta_{k}}=-(X^{T}y)_{k}+(X^{T}X\beta)_{k}+\lambda_{1}s_{k}+\lambda_{2}\sum_{l:(k,l)\in E}t_{kl}\label{eq:GeneralFusedLassoSubgrad}
\end{equation}
where $s_{k}=\mbox{sign}(\beta_{k})$ for $\beta_{k}\neq0$ and $s_{k}\in[-1,1]$
otherwise. Also, as before $t_{kl}=\mbox{sign}(\beta_{k}-\beta_{l})$
for $\beta_{k}\neq\beta_{l}$ and $t_{kl}\in[-1,1]$ for $\beta_{k}=\beta_{l}$.
Variables are considered fused if $\beta_{k}=\beta_{l}$
and unfused if $\beta_{k}\neq\beta_{l}$. They are active if
$\beta_{k}\neq0$, from which we see that in a set of fused variables, always all variables 
are either active or not. Thus, when ``activating''
coefficients, we always have to activate a whole set at once.
In order to do this, we also have to keep track of sets of variables
that are inactive by using $s_{k}$ as a substitute for $\beta_{k}$.
More precisely:

\begin{definition}{3}\label{def:FusedActiveSets} 
Let $p_{F}(\lambda_{2})$
be the number of sets of fused variables for penalty parameter $\lambda_{2}$
(keeping $\lambda_{1}$ fixed). Then for the sets $F_{i}$, $i=1,\ldots,p_{F}(\lambda_{2})$
to be valid, the following conditions have to hold:

\begin{enumerate}
\item $\cup_{i=1}^{n_{F}(\lambda_{2})}F_{i}=\{1,\ldots,n\}$
\item $F_{i}\cap F_{j}=\emptyset \quad \mbox{for} \quad i\neq j$ 
\item If $k,l\in F_{i}$ then $\beta_{k}(\lambda_{2})=\beta_{l}(\lambda_{2})$
as well as $s_{k}(\lambda_{2})=s_{l}(\lambda_{2})$. If $k\in F_{i,}l\in F_{j},i\neq j$
and $F_{i}$ and $F_{j}$ have a connecting edge, then $\beta_{k}(\lambda_{2})\neq\beta_{l}(\lambda_{2})$
or $s_{k}(\lambda_{2})\neq s_{l}(\lambda_{2})$ for all penalty parameters
in an interval $(\lambda_{2},\lambda_{2}+\varepsilon)$ for some $\varepsilon>0$.
\item If $k,l\in F_{i}$ then $k$ and $l$ are connected in $\mathcal{G}$
by only going over nodes in $F_{i}$, i.e. $k,l$ are connected in
$\mathcal{G}|_{F_{i}}$, the subgraph of $\mathcal{G}$ induced by
$F_{i}$.
\end{enumerate}
\end{definition}

This definition is almost the same as Definition 2
except for a small change in point 3. This change reflects the adjustment
for inactive coefficients as mentioned above. In addition to this,
we also need a more formal definition of what active and inactive
sets are. 

\begin{definition}{4}\label{def:ActiveSets}For penalty parameter $\lambda_{2}^{0}$
(fixing $\lambda_{1}$), let $\mathcal{A}(\lambda_{2}^{0})$ be the
set of active fused sets and $\mathcal{N}(\lambda_{2}^{0})$ be the
set of inactive fused sets. Then 
\[
\mathcal{A}(\lambda_{2}^{0})=\left\{ i|\beta_{F_{i}}(\lambda)\neq0\,\mbox{for}\,\lambda\in(\lambda_{2}^{0},\lambda_{2}^{0}+\varepsilon)\right\} 
\]
and
\[
\mathcal{N}(\lambda_{2}^{0})=\left\{ 1,\ldots,p_{F}(\lambda_{2}^{0})\right\} \backslash\mathcal{A}(\lambda_{2}^{0}).
\]
\end{definition}

Here, a set is defined as active at $\lambda_2^0$ not if $\beta_{F_i}(\lambda_2^0)\neq 0$ 
but instead if $\beta_{F_i}(\lambda_2)\neq 0$ for $\lambda_2 > \lambda_2^0$. There is only a difference
between these two definitions at the breakpoints of the piecewise linear path. As we increase $\lambda$ from $0$ to $\infty$, the definition for active sets was chosen to be ``forward looking''. This distinction is of a technical nature that is mostly important for the algorithm and proofs later. 

Now that we have defined fused and active sets, we will determine
how $\beta_{k}$ as well as $s_{k}$ and $t_{kl}$ change with $\lambda_{2}$.

\subsection{Derivation of the algorithm }
\subsubsection{The loss function}

The derivation of the algorithm will work similar to the sections
before. First, we will incorporate the conditions of fused and active
sets into the loss function and then find the derivative of $\beta_{F_{i}}(\lambda_{2})$
with respect to $\lambda_{2}$. After this, we will determine how
$s_{k}$ and $t_{kl}$ change with $\lambda_{2}$ and use it to specify
when sets of variables are becoming active or inactive and have to
be fused or split. Afterwards, we will prove that the solution is
piecewise linear in $\lambda_{2}$ and assemble everything into a
path algorithm. 

Before going into more details, we need to define some additional
notation. Using the sets of fused variables and active sets, we specify
a predictor matrix that incorporates this information. Using the sets
of fused variables $F_{i}$ for $i=1,\ldots,p_{F}(\lambda_{2})$ define
$\bX^{F}\in \mathds{R}^{n\times p_{F}(\lambda_{2})}$ with
\[
\bx_{i}^{F}=\sum_{k\in F_{i}}\bx_{k}
\]
and based on this we incorporate the active sets into 
$\bX^{F,\mathcal{A}}\in \mathds{R}^{n\times|\mathcal{A}(\lambda_{2})|}$
by dropping all columns in $\bX^{F}$ that do not correspond to active
sets. Similarly, by $\bX^{\mathcal{A}}$ we refer to the columns of
$\bX$ corresponding to coefficients in active sets, i.e. $k\in F_{i}$
with $i\in\mathcal{A}(\lambda_{2})$, and by $\bX^{\mathcal{N}}$ to
the columns of coefficients in inactive sets. Then the constrained
loss function incorporating active and fused sets is 
\begin{align}
L_{F,\mathcal{A},\lambda_{1},\lambda_{2}}(\by,\bX,\bbeta) & =\frac{1}{2}(\by-\bX^{F}\bbeta^{F})^{T}(\by-\bX^{F}\bbeta^{F})+\lambda_{1}\sum_{i\in\mathcal{A}(\lambda_{2})}|F_{i}||\beta_{i}^{F}|+\label{eq:generalFusedLassoConst}\\
& +\lambda_{2}\sum_{i<j}\left|\left\{ (k,l):k\in F_{i,}l\in F_{j}\right\} \right|\left|\beta_{i}^{F}-\beta_{j}^{F}\right|
\end{align}
where $\bbeta^{F}$ is a vector such that $\beta_{i}^{F}=\beta_{F_{i}}$.
Assuming that the fused and active sets are correct, the minimizer
of $L_{F,\mathcal{A},\lambda_{1},\lambda_{2}}$ is also the minimizer
of the original loss function $L_{\lambda_{1},\lambda_{2}}$. Note
that by definition of the fused sets $\beta_{F_{i}}\neq\beta_{F_{j}}$
(except for a finite number of breakpoints). Therefore, the subgradient
equations for this constrained loss function with respect to $\lambda_{2}$
are
\begin{eqnarray*}
\frac{\partial L_{F,\mathcal{A},\lambda_{1},\lambda_{2}}(\by,\bX,\bbeta)}{\partial\beta_{i}^{F}} & = & -\bX_{i}^{F,T}\by+\bX_{i}^{F,T}\bX^{F}\bbeta^{F}+\lambda_{1}|F_{i}|s_{F_{i}}+\\
& + & \lambda_{2}\sum_{j\in\mathcal{A}(\lambda_{2})}\left|\left\{ (k,l):k\in F_{i,}l\in F_{j}\right\}
\right|\mbox{sign}\left(\beta_{F_{i}}-\beta_{F_{j}}\right)=0,
\end{eqnarray*}
where $s_{F_{i}}$ is a short form for $s_{k}$ with $k\in F_{i}$
just as $\beta_{F_{i}}$. Now, if we define vectors $\ba$ and $\bb$
with 
\[
a_{i}=|F_{i}|s_{F_{i}}\,\,\,\,\,\mbox{and}\,\,\,\,\, b_{i}=\sum_{j\neq i}\left|\left\{ (k,l):k\in F_{i,}l\in F_{j}\right\} \right|\mbox{sign}\left(\beta_{F_{i}}-\beta_{F_{j}}\right)
\]
then we can write this optimality condition more compact as\begin{eqnarray*}
\frac{\partial L_{F,\mathcal{A},\lambda_{2}}(\by,\bX,\bbeta)}{\partial\bbeta^{F}} & = & -\bX^{F,T}\by+\bX^{F,T}\bX^{F}\bbeta^{F}+\lambda_{1}\ba+\lambda_{2}\bb=0.\end{eqnarray*}

For $i\in\mathcal{A}(\lambda_{2})$, we know that $s_{i}^{F}(\lambda_{2})=s_{F_{i}}(\lambda_{2})=1$
locally w.r.t. $\lambda_{2}$ and for $i\in\mathcal{N}(\lambda_{2})$
we have $\beta_{i}^{F}(\lambda_{2})=\beta_{F_{i}}(\lambda_{2})=0$.
Therefore, when taking derivatives with respect to $\lambda_{2}$
we get (splitting into active and inactive sets)

\[
(\bX^{F,\mathcal{A}})^{T}\bX^{F,\mathcal{A}}\frac{\partial\bbeta^{F,\mathcal{A}}}{\partial\lambda_{2}}+\bb^{\mathcal{A}}=0
\]
as well as
\[
(\bX^{F,\mathcal{N}})^{T}\bX^{F,\mathcal{A}}\frac{\partial\bbeta^{F,\mathcal{A}}}{\partial\lambda_{2}}+\lambda_{1}\frac{\partial \ba^{\mathcal{N}}}{\partial\lambda_{2}}+\bb^{\mathcal{N}}=0.
\]

We can solve this by
\begin{equation}
\frac{\partial\bbeta^{F,\mathcal{A}}}{\partial\lambda_{2}}=-\left((\bX^{F,\mathcal{A}})^{T}\bX^{F,\mathcal{A}}\right)^{-1}\bb^{\mathcal{A}}\label{eq:BetaDerivGeneral}
\end{equation}
and
\begin{equation}
\frac{\partial \ba^{\mathcal{N}}}{\partial\lambda_{2}}=-\frac{1}{\lambda_{1}}\left((\bX^{F,\mathcal{N}})^{T}\bX^{F,\mathcal{A}}\frac{\partial\bbeta^{F,\mathcal{A}}}{\partial\lambda_{2}}+\bb^{\mathcal{N}}\right)\label{eq:SDerivGeneral}
\end{equation}
from which we get $\partial s_{F_{i}}/\partial\lambda_{2}=(1/|F_{i}|) \cdot (\partial a_{i}^{\mathcal{N}}/\partial\lambda_{2})$. Here, by the condition that $\text{rank}(X)=p$, it is guaranteed that $\left((\bX^{F,\mathcal{A}})^{T}\bX^{F,\mathcal{A}}\right)^{-1}$ exists.
Therefore, we have calculated the derivative of $\bbeta^{F}$ and $\bs^{F}$
from Equation (\ref{eq:generalFusedLassoConst}). From this, we can
derive the slope of the path for $\beta^{F}$ as well as determine
when sets of coefficients become active or inactive. However, in order
to see when to split a set, it remains to determine the behavior of
$t_{kl}$ with respect to $\lambda_{2}$ or equivalently of $\tau_{kl}=\lambda_{2}t_{kl}$.
We do this by inserting the solutions found above into the subgradient
equations of the unconstrained problem (\ref{eq:GeneralFusedLassoSubgrad}).
As before, we assume that we have a solution at $\lambda_{2}$ that
satisfies the subgradient equations. Then for any coefficient $k\in F_{i}$
we have
\[
-\bx_{k}^{T}\by+\bx_{k}^{T}\bX^{\mathcal{A}}\bbeta^{\mathcal{A}}+\lambda_{1}s_{k}+\lambda_{2}\sum_{l:(k,l)\in E;k,l\in F_{i}}t_{kl}+\lambda_{2}\sum_{l:(k,l)\in E;l\not\in F_{i}}t_{kl}=0,
\]
again grouping the $t_{kl}$ by whether $k$ and $l$ belong to the
same or different sets. When taking the derivative with respect to
$\lambda_{2}$, we get
\begin{equation}
\bx_{k}^{T}\bX^{\mathcal{A}}\frac{\partial\bbeta^{\mathcal{A}}}{\partial\lambda_{2}}+\lambda_{1}\frac{\partial s_{k}}{\partial\lambda_{2}}+\sum_{l:(k,l)\in E;k,l\in F_{i}}\frac{\partial\tau_{kl}}{\partial\lambda_{2}}+\sum_{l:(k,l)\in E;l\not\in F_{i}}t_{kl}=0\label{eq:subgradientGeneralFusedLasso}
\end{equation}

In order to determine $\partial\tau_{kl}/\partial\lambda_{2}$,
we use the maximum flow setup from the general FLSA algorithm as well
as that $t_{kl}=\mbox{sign}(\beta_{k}-\beta_{l})$ for $k\in F_{i},l\in F_{i}$.
Then we define the push $p_{k}$ on node $k$ as 
\[
p_{k}=-\bx_{k}^{T}\bX^{\mathcal{A}}\frac{\partial\bbeta^{\mathcal{A}}}{\partial\lambda_{2}}-\lambda_{1}\frac{\partial s_{k}}{\partial\lambda_{2}}-\sum_{l:(k,l)\in E;l\not\in F_{i}}t_{kl}
\]
and therefore we have to solve 
\[
p_{k}=\sum_{l:(k,l)\in E;k,l\in F_{i}}\frac{\partial\tau_{kl}}{\partial\lambda_{2}}\,\,\,\,\,\mbox{for}\,\,\,\,\, k=1,\ldots,p\]
This is exactly the same problem as in Section 3.2.2
and we use the same maximum flow setup described there to find $\partial\tau_{kl}/\partial\lambda_{2}$
by setting $\partial\tau_{kl}/\partial\lambda_{2}=f_{kl}$
where $f_{kl}$ is the maximal flow from node $k$ to node $l$ in
$F_{i}$. Using all this, we can now show that the solution is piecewise
linear in $\lambda_{2}$ in the following theorem:

\begin{thm}{4}\label{ThmSolutionLinearGeneral}For some $\lambda_{2}^{0}$,
let $F_{1},\ldots,F_{p_{F}(\lambda_{2}^{0})}$ be a valid grouping
of the variables. Let $\tilde{\mathcal{G}_{i}}=(\tilde{V}_{i},\tilde{E}_{i},\tilde{C}_{i})$
be the with $F_{i}$ associated maximum-flow graph as defined above.
Also let $\beta_{k}(\lambda_{2}^{0}),s_{k}(\lambda_{2}^{0})$ and
$\tau_{kl}(\lambda_{2}^{0})$ be a solution to the Fused Lasso problem
for penalty parameter $\lambda_{2}=\lambda_{2}^{0}$. Assume that $\tilde{\mathcal{G}_{i}}$
has a maximum flow for which all flows coming from the source are
at maximum capacity (i.e. $f_{rl}=c_{rl}$ for all $(r,l)\in\tilde{E}_{i}$),
and $\partial\beta_{F_{i}}/\partial \lambda_{2}$ is as defined in Equation
(\ref{eq:BetaDerivGeneral}) as well as $\partial s_{k}/\partial\lambda_{2}$
as in (\ref{eq:SDerivGeneral}). Then there exists some $\Delta>0$
such that for any $\lambda_{2}\in[\lambda_{2}^{0},\lambda_{2}^{0}+\Delta]$,
the solution to the Fused Lasso problem is given by
\begin{eqnarray*}
\beta_{k}(\lambda_{2}) & = & \beta_{k}(\lambda_{2}^{0})+\frac{\partial\beta_{F_{i}}}{\partial \lambda_{2}}(\lambda_2^0)\cdot (\lambda_{2}-\lambda_{2}^{0})\quad\mbox{for}\quad k\in F_{i}\\
s_{k}(\lambda_{2}) & = & s_{k}(\lambda_{2}^{0})+\frac{\partial s_{F_{i}}}{\partial \lambda_{2}}(\lambda_2^0)\cdot (\lambda_{2}-\lambda_{2}^{0})\quad\mbox{for}\quad k\in F_{i}
\end{eqnarray*}

and
\begin{eqnarray*}
\tau_{kl}(\lambda_{2}) & = & \begin{cases}
\tau_{kl}(\lambda_{2}^{0})+f_{kl}(\lambda_{2}-\lambda_{2}^{0}) & \mbox{for }k,l\in F_{i}\quad\mbox{for some }i\\
\mbox{sign}(\tau_{kl}(\lambda_{2}^{0}))\lambda_{2} & \mbox{otherwise.}\end{cases}
\end{eqnarray*}
\end{thm}

The proof is very similar to the one for Theorem 3
and again can be found in Section \ref{AppProofSolutionLinearGeneral}. As before, we still have to define
what $\Delta$ is.

\subsubsection{Activation and deactivation time}

In order to define $\Delta$, we have to find the value of $\lambda_{2}$
at which the fused sets themselves or their activation status changes. The hitting
time $h(\lambda_{2})$ and violation time $v(\lambda_{2})$ are the
same as in Section 3.2.4. In addition to this
we also have to define the value of $\lambda_{2}$ at which sets are
activated or deactivated. The activation time $act(\lambda_{2})$
is 
\[
act(\lambda_{2})=\min_{i\in\mathcal{N}(\lambda_{2})}act_{i}(\lambda_{2})
\]
where
\[
act_{i}(\lambda_{2})=\begin{cases}
\frac{1-s_{F_{i}}}{\frac{\partial s_{F_{i}}}{\partial\lambda_{2}}}+\lambda_{2} & \mbox{if}\,\,\frac{\partial s_{F_{i}}}{\partial\lambda_{2}}>0\\
\frac{s_{F_{i}}+1}{-\frac{\partial s_{F_{i}}}{\partial\lambda_{2}}}+\lambda_{2} & \mbox{if}\,\,\frac{\partial s_{F_{i}}}{\partial\lambda_{2}}<0\\
\infty & \mbox{otherwise}\end{cases}
\]
is the activation time of set $F_{i}$ with $i\in\mathcal{N}(\lambda_{2})$.
The deactivation time on the other hand $d(\lambda_{2})$ is
\[
d(\lambda_{2})=\min_{i\in\mathcal{A}(\lambda_{2})}d_{i}(\lambda_{2})
\]
where 
\[
d_{i}(\lambda_{2})=\begin{cases}
\frac{\beta_{F_{i}}}{-\frac{\partial\beta_{F_{i}}}{\partial\lambda_{2}}}+\lambda_{2} & \mbox{if}\,\,\frac{\partial\beta_{F_{i}}}{\partial\lambda_{2}}<0,\beta_{F_{i}}>0\\
\frac{-\beta_{F_{i}}}{\frac{\partial\beta_{F_{i}}}{\partial\lambda_{2}}}+\lambda_{2} & \mbox{if}\,\,\frac{\partial\beta_{F_{i}}}{\partial\lambda_{2}}>0,\beta_{F_{i}}<0\\
\infty & \mbox{otherwise}\end{cases}
\]
is the deactivation time of the active set $F_{i}$. Putting all this
together, the length of the linear segment to the next breakpoint
is
\[
\Delta(\lambda_{2})=\min\left\{ h(\lambda_{2}),v(\lambda_{2}),act(\lambda_{2}),d(\lambda_{2})\right\} -\lambda_{2}.
\]

Therefore, we have now defined all the necessary information on the previous theorem.

\subsubsection{Changing sets of variables}

What remains to be done is to find the rule on how to split and fuse
sets of variables as well as how to activate or inactivate a set.
First, the rule for splitting and fusing sets is very similar to the
one in Section 3.2.4 with the only change being
in step one to take the $s_{k}$ into account so that inactive sets
can also be fused.

\begin{enumerate}
\item If there are sets $F_{i}$ and $F_{j}$ for which $\exists k\in F_{i}$
and $l\in F_{j}$ with $(k,l)\in E$ and $\beta_{F_{i}}=\beta_{F_{j}}$
and $s_{F_{i}}=s_{F_{j}}$, then fuse these sets into a new set $\tilde{F}_{ij}=F_{i}\cup F_{j}$
if $\frac{\partial\beta_{F_{i}}}{\partial\lambda_{2}}-\frac{\partial\beta_{F_{j}}}{\partial\lambda_{2}}\le0$,
$\frac{\partial s_{F_{i}}}{\partial\lambda_{2}}-\frac{\partial s_{F_{j}}}{\partial\lambda_{2}}\le0$
and $t_{kl}=1$.
\item If there is a set $F_{i}$ for which in the associated maximal flow
graph not all edges coming from the source are at maximal capacity,
then split $F_{i}$ in the two subsets $R_{i}$ and $S_{i}$ as described
above. 
\item Iterate steps 2 and 3 until nothing changes. 
\end{enumerate}
Second, the rule for activating and inactivating the sets is then:

\begin{enumerate}
\item If for an inactive set $F_{i}$ we have $\frac{\partial s_{F_{i}}}{\partial\lambda_{2}}>0$
if $s_{F_{i}}=1$ or $\frac{\partial s_{F_{i}}}{\partial\lambda_{2}}<0$
if $s_{F_{i}}=-1$, then activate set $F_{i}$.
\item If for an active set $F_{i}$ we have $\beta_{F_{i}}=0$ and $s_{F_{i}}=1$,
$\frac{\partial\beta_{F_{i}}}{\partial\lambda_{2}}<0$ or $s_{F_{i}}=-1$,
$\frac{\partial\beta_{F_{i}}}{\partial\lambda_{2}}>0$, then deactivate
the set.
\end{enumerate}
When we use these two rules for values of $\lambda_{2}$ at which
we have to adapt the fused and active sets, then the resulting sets
will be valid. More precisely, the following proposition will hold:

\begin{prop}{2}\label{prop:ChangeVariablesFusedLasso}Assume that sets
of variables $F_{i}$, $i=1,\ldots,p_{F}(\lambda_{2}-)$ are given
(where $\lambda_{2}-$ denotes a limit from the left) with optimal
values of $\beta_{F_{i}}(\lambda_{2})$, $s_{F_{i}}(\lambda_{2})$
and $t_{kl}(\lambda_{2})$ for all $k,l$. Furthermore assume that
$\partial\beta_{F_{i}}/\partial\lambda_{2}$ and $\partial s_{F_{i}}/\partial\lambda_{2}$
are calculated as defined in equations (\ref{eq:BetaDerivGeneral})
and (\ref{eq:SDerivGeneral}) using the sets $F_{i}$. When using the
two rules above for changing the fused sets and the activity status,
the resulting sets will be valid according to Definition \ref{def:FusedActiveSets}
and \ref{def:ActiveSets}.\end{prop}

The proof can be found in Section \ref{AppProofChangeVariablesFusedLasso}. This proposition
completes the necessary steps for the path algorithm and we can now
present an outline of the whole algorithm.

\subsection{Outline of the algorithm}

The algorithm is an extension of the FLSA algorithm that incorporates
the active and inactive sets. In the case of the general FLSA algorithm,
finding the starting value for $\lambda_{2}=0$ was particularly easy.
For general Fused Lasso, finding the starting value needs an additional
step. For $\lambda_{2}=0$, the loss function is
\[
L_{\lambda_{1},0}(\by,\bX,\bbeta)=\frac{1}{2}(\by-\bX\bbeta)^{T}(\by-\bX\bbeta)+\lambda_{1}\sum_{k=1}^{p}|\beta_{k}|
\]
which is just the loss function of the regular Lasso. Therefore, we
can find the starting values for $\beta$ by solving a regular Lasso
problem. In order to find the starting value for $s$, we have to
look at the subgradient equations:
\[
\frac{\partial L_{\lambda_{1},\lambda_{2}}}{\partial\beta_{k}}=-(\bX^{T}\by)_{k}+(\bX^{T}\bX\bbeta)_{k}+\lambda_{1}s_{k}=0
\]
and thus
\[
s=\frac{1}{\lambda_{1}}(\bX^{T}\by-\bX^{T}\bX\bbeta).
\]

Using these starting values we now have the complete algorithm which can be seen in Algorithm \ref{GeneralFusedLassoAlg}.

\begin{algorithm}[tbp]
\caption{General Fused Lasso path algorithm}
\SetLine
\SetKwBlock{Initialize}{initialize}{end}
\Initialize{$\lambda_{2}=0$\;
$F_{i}=\{i\}$ for $i=1,\ldots,p$\;
$p_{F}=p$\;
Find $\beta_{k}$ and $s_{k}$ for $k=1,\ldots,n$ by solving the
Lasso problem\;
}%
\While{$p_{F}>1$}{Update $\beta$, $s$ and $t$ \;
Calculate the derivatives of $\beta_{F_{i}}$ and $s_{F_{i}}$ w.r.t.
$\lambda_{2}$ for $i=1,\ldots,p_{F}$\;
Solve maximum flow problem for $F_{i}$, $i=1,\ldots,p_{F}(\lambda_{2})$\;
Calculate the next hitting time $h(\lambda_{2})$\;
Calculate the next violation time $v(\lambda_{2})$\;
Calculate the next time a set will be activated $act(\lambda_{2})$
or deactivated $d(\lambda_{2})$\;
Set $\Delta(\lambda_{2})=\min\left\{ h(\lambda_{2}),v(\lambda_{2}),a(\lambda_{2}),d(\lambda_{2})\right\} -\lambda_{2}$\;
\uIf{$h_{ij}(\lambda_{2})=\Delta(\lambda_{2})+\lambda_{2}$}{Fuse
the two sets $F_{i}$ and $F_{j}$\;$p_{F}:=p_{F}-1$\;$\lambda_{2}:=h_{ij}(\lambda_{2})$\;}\uElseIf{$v_{i}(\lambda_{2})=\Delta(\lambda_{2})+\lambda_{2}$}{\If{$v_{i}(\lambda_{2})=\lambda_{2}$}{Split
the set $F_{i}$ into two smaller sets\; $p_{F}:=p_{F}+1$ \; }
$\lambda_{2}:=v_{i}(\lambda_{2})$ \; }\uElseIf{ $a_{i}(\lambda_{2})=\Delta(\lambda_{2})+\lambda_{2}$
}{Activate the set $F_{i}$\; $\lambda_{2}:=a_{i}(\lambda_{2})$
\;}\uElseIf{$d_{i}(\lambda_{2})=\Delta(\lambda_{2})+\lambda_{2}$}{Deactivate
the set $F_{i}$ \; $\lambda_{2}:=d_{i}(\lambda_{2})$ \;}}\label{GeneralFusedLassoAlg}
\end{algorithm}

Most of the components of this algorithm are very similar to the FLSA
algorithm. However, many efficiency gains that we could get in the
FLSA case are not possible for the general Fused Lasso. For the FLSA,
the derivative of $\beta_{F_{i}}$ and $t_{kl}$ for $k,l\in F_{i}$
only had to be updated if $F_{i}$ was the result of the last split
or fusion. For the Fused Lasso on the other hand, if only one of the
active groups changes its composition, the path for all other groups
are affected as well and have to be recomputed, including the maximum
flow problems. Only inactive groups can be treated as in the FLSA
case. Another computational bottleneck is the calculation of $\left((\bX^{F,\mathcal{A}})^{T}\bX^{F,\mathcal{A}}\right)^{-1}$.
It is worthwhile to note that it is not necessary to recompute the
whole matrix in every step. As from one step to the next, only very
few columns of $\bX^{F,\mathcal{A}}$ change, it is possible to get
the new matrix inverse by up- and downdating the old matrix inverse. 

Overall, the General Fused Lasso algorithm is computationally a lot
more complex than the FLSA case. Apart from this, the more immediate
applications are for the FLSA, especially the one-dimensional case.
Therefore, we will only publish implementations for the FLSA cases
at the moment and delay a program that solves the general version
of the Fused Lasso to a later date.

\subsection{Proof of Theorem \ref{ThmSolutionLinearGeneral}}
\label{AppProofSolutionLinearGeneral}
\begin{proof}By the assumption, the values for $\beta_{F_{i}}(\lambda_{2}^{0})$,
$s_{F_{i}}(\lambda_{2}^{0})$ and $t_{kl}(\lambda_{2}^{0})$ satisfy
the subgradient equations (\ref{eq:GeneralFusedLassoSubgrad}) for
the value $\lambda_{2}=\lambda_{2}^{0}$. Then the subgradient equations
for value $\lambda_{2}\in[\lambda_{2}^{0},\lambda_{2}^{0}+\Delta]$
are \begin{align*}
&-(\bX^{T}\by)_{k}+(\bX^{T}\bX\bbeta(\lambda_{2}))_{k}+\lambda_{1}s_{k}(\lambda_{2})+ \lambda_{2}\sum_{l:(k,l)\in E}t_{kl}(\lambda_{2})  \\
&=-\bx_{k}^{T}\by+\bx_{k}^{T}\bX\bbeta(\lambda_{2}^{0})+\lambda_{1}s_{k}(\lambda_{2}^{0})+
\sum_{l:(k,l)\in E,l\not\in F_{i}}\tau_{kl}(\lambda_{2}^{0})+\sum_{l:(k,l)\in E,l\in F_{i}}\tau_{kl}(\lambda_{2}^{0})  \\
&+(\lambda_{2}-\lambda_{2}^{0})\left(\bx_{k}^{T}\bX\frac{\partial\bbeta}{\partial\lambda_{2}}+\lambda_{1}\frac{\partial s_{k}}{\partial\lambda_{2}}+\sum_{l:(k,l)\in E,l\not\in F_{i}}t_{kl}(\lambda_2^0)+\sum_{l:(k,l)\in E,l\in F_{i}}f_{kl}\right) \\
&=0\end{align*}
and we have to show that they are equal to zero. As by assumption $\bbeta(\lambda_2^0)$, $\bs(\lambda_2^0)$ and $\btau(\lambda_2^0)$ are by assumption optimal, it suffices to show that for all $k$ 
\[
\bx_{k}^{T}\bX\frac{\partial\bbeta}{\partial\lambda_{2}}+\lambda_{1}\frac{\partial s_{k}}{\partial\lambda_{2}}+\sum_{l:(k,l)\in E,l\not\in F_{i}}t_{kl}(\lambda_{2}^{0})+\sum_{l:(k,l)\in E,l\in F_{i}}f_{kl}=0.
\]
By the definition of the push on node $k$, this in
turn is equivalent to 
\[
-p_{k}+\sum_{l:(k,l)\in E,l\in F_{i}}f_{kl}=0.
\]

In order to see this, observe that by assumption in the theorem, all nodes coming from the source are
at maximum capacity. Furthermore, note that as $t_{kl}=-t_{lk}$,
we also have that 
\begin{align*}
\sum_{k\in F_{i}}p_{k}&=-(\bX_{i}^{F})^{T}\bX^{\mathcal{A}}\frac{\partial\bbeta^{\mathcal{A}}}{\partial\lambda_{2}}-\lambda_{1}|F_{i}|\frac{\partial s_{F_{i}}}{\partial\lambda_{2}}\\
&-\sum_{j\neq i}\left|\left\{ (k,l):k\in F_{i,}l\in F_{j}\right\} \right|\mbox{sign}\left(\beta_{F_{i}}-\beta_{F_{j}}\right) =0
\end{align*}
as can be seen in the derivation of the algorithm. Therefore, 
\[
\sum_{k\in F_{i},p_{k}>0}p_{k}=-\sum_{k\in F_{i},p_{k}<0}p_{k}
\]
and thus, if all flows coming from the source are at capacity, so
are all flows going into the sink. As the flows going into and out
of every regular node $k$ in the maximum flow problem sum up to 0,
this just implies that for every node $k$ \[
-p_{k}+\sum_{l:(k,l)\in E,l\in F_{i}}f_{kl}=0\]
holds. And this finishes the proof of the theorem. \end{proof}

\subsection{Proof of Proposition \ref{prop:ChangeVariablesFusedLasso}}
\label{AppProofChangeVariablesFusedLasso}

\begin{proof}
This proof is very similar to the proof of Proposition
2 above. 

Before we start with the main part, note first that $\bbeta(\lambda_2)$ is a continuous function of $\lambda_2$. Under the condition that $\text{rank}(X)=p$, the loss function $L_{\lambda_{1},\lambda_{2}}(\by,\bX,\bbeta)$ is strictly convex in $\bbeta$ and affine in $\lambda_2$. From this we can immediately see that
$
\min_\bbeta L_{\lambda_{1},\lambda_{2}}(\by,\bX,\bbeta)
$
is a continuous function of $\lambda_2$. However, then the strict convexity in $\bbeta$ of $L$ implies the continuity of the solution $\bbeta(\lambda_2)$. 

Now back to the actual proof. In order for a set $F_{i}$ to be
valid at $\lambda_{2}^{0}$, it has to hold that for any other set
$F_{j}$ that $\beta_{F_{i}}(\lambda_{2})\neq\beta_{F_{j}}(\lambda_{2})$
or $s_{F_{i}}(\lambda_{2})\neq s_{F_{j}}(\lambda_{2})$ for $\lambda_{2}\in(\lambda_{2}^{0},\lambda_{2}^{0}+\varepsilon)$
for some $\varepsilon>0$. 

If $\beta_{F_{i}}(\lambda_{2}^{0})\neq\beta_{F_{j}}(\lambda_{2}^{0})$
or $s_{F_{i}}(\lambda_{2}^{0})\neq s_{F_{j}}(\lambda_{2}^{0})$, then
this clearly holds as $\beta$ and $s$ are continuous in $\lambda_{2}$.

If $\beta_{F_{i}}=\beta_{F_{j}}$ and $s_{F_{i}}=s_{F_{j}}$, then
the two sets will be fused unless one of the conditions $\partial\beta_{F_{i}}/\partial\lambda_{2}-\partial\beta_{F_{j}}/\partial\lambda_{2}\le0$,
$\partial s_{F_{i}}/\partial\lambda_{2}-\partial s_{F_{j}}/\partial\lambda_{2}\le0$
or $t_{kl}=1$ is violated. First, as $F_{i}$ and $F_{j}$ are different
sets, we have that $t_{kl}=\pm1$ and as $t_{kl}=-t_{lk}$ by definition,
we can assume without loss of generality that $t_{kl}=1$. If either
one of the other two conditions fails, then either $\beta_{F_{i}}(\lambda_{2})>\beta_{F_{j}}(\lambda_{2})$
or $s_{F_{i}}(\lambda_{2})>s_{F_{j}}(\lambda_{2})$ for $\lambda_{2}\in(\lambda_{2}^{0},\lambda_{2}^{0}+\varepsilon)$
and it is also not violating the restriction that $t_{kl}=\mbox{sign}(\beta_{k}-\beta_{l})$
for $\beta_{k}\neq\beta_{l}$ where $(k,l)\in E$ with $k\in F_{i}$
and $l\in F_{j}$. Therefore, the set is valid. 

Another concern that we have to address is that we have to ensure the existence of $t_{kl}$ that satisfy the constraints $t_{kl} \in [-1,1]$ for $k,l\in F_i$ for all $i$. However, if a flow exists in $\tilde{\mathcal{G}}_i$, by the construction of the flow graph, this holds. 

Therefore, for the fusing and splitting steps, it remains to show that there cannot be an
infinite loop of fusions and splits. As there is only a
finite number of nodes, it is enough to show that if two sets $F_{i}$
and $F_{j}$ have just been merged, they cannot be split again into
$F_{i}$ and $F_{j}$ immediately at $\lambda_{2}=\lambda_{2}^{0}$.

In order to do this, assume that there is
only one breakpoint exactly at $\lambda_{2}=\lambda_{2}^{0}$. Otherwise
let $\be\sim N(0,I_{n})$ and perturb $\by$ by using $\by+c\be$ where $c>0$ and send $c \rightarrow 0$. The probability of two breakpoints
occurring for the same $\lambda_{2}$ for the random $\by+c\be$ is $0$
and the solutions are all linear in $c$, thus the solution if just $\by$ is being used is equal to the limit of the solution using $\by+c\be$ for $c \rightarrow 0$.
 
Therefore, we can without loss of generality assume that there is only one breakpoint. Also,
we assume that $\beta_{F_i}(\lambda_2^0)=\beta_{F_j}(\lambda_2^0)$ as well as $\beta_{F_{i}}(\lambda_{2})>\beta_{F_{j}}(\lambda_{2})$ for $\lambda_{2}\in(\lambda_{2}^{0}-\varepsilon,\lambda_{2}^{0})$
and therefore $\partial \beta_{F_{i}}/\partial\lambda_{2} < \partial \beta_{F_{j} }/ \partial \lambda_{2}$ (the same remains true when we use $\bs$ instead of $\bbeta$). This implies that the two groups $F_{i}$ and $F_{j}$ have to remain fused for at least an interval $[\lambda_{2}^{0},\lambda_{2}^{1}]$
for $\lambda_{2}^{0}<\lambda_{2}^{1}$. To see this, observe that
there are only two other possible cases:

\begin{description}
\item [{$\beta_{F_{i}}(\lambda_{2})>\beta_{F_{j}}(\lambda_{2})\,\,\mbox{for}\,\,\lambda_{2}>\lambda_{2}^{0}$}:] This
would require that $\partial \beta_{F_{i}}/\partial \lambda_{2}(\lambda_{2}) > \partial \beta_{F_{j}}/ \partial \lambda_{2}(\lambda_{2})$
as $\beta_{F_{i}}(\lambda_{2}^{0}) = \beta_{F_{j}}(\lambda_{2}^{0})$. However, this is not
possible as in this case, equation (\ref{eq:BetaDerivGeneral}) did
not change compared to $\lambda_{2}<\lambda_{2}^{0}$. Therefore
$\partial \beta_{F_{i}}/\partial\lambda_{2}<\partial \beta_{F_{j}}/\partial\lambda_{2}$
as before, contradicting $\beta_{F_{i}}(\lambda_{2})>\beta_{F_{j}}(\lambda_{2})$ for $\lambda_2 > \lambda_2^0$.
\item [{$\beta_{F_{i}}(\lambda_{2})<\beta_{F_{j}}(\lambda_{2})\,\,\mbox{for}\,\,\lambda_{2}>\lambda_{2}^{0}$}:] In
this case we would have that due to the continuity of $\beta$ it
holds
\begin{align*}
\lim_{\lambda_{2}\uparrow\lambda_{2}^{0}}\frac{\partial L_{F,\mathcal{A},\lambda_{1},\lambda_{2}}(\by,\bX,\bbeta)}{\partial\beta_{i}^{F}}
=\lim_{\lambda_{2}\downarrow\lambda_{2}^{0}}\frac{\partial L_{F,\mathcal{A},\lambda_{1},\lambda_{2}}(\by,\bX,\bbeta)}{\partial\beta_{i}^{F}}+2\lambda_{2}\left|\left\{ (k,l):k\in F_{i,}l\in F_{j}\right\} \right|.
\end{align*}
By assumption we know that $\bbeta(\lambda_2)$ is optimal for $\lambda_2 < \lambda_2^0$ and thus 
$$
L_{F,\mathcal{A},\lambda_{1},\lambda_{2}}(\by,\bX,\bbeta) = 0 \quad \text{for} \quad \lambda_2 < \lambda_2^0.
$$
However this implies then that $\bbeta(\lambda_2)$ for which $\beta_{F_{i}}(\lambda_{2})<\beta_{F_{j}}(\lambda_{2})$ can not be optimal for $\lambda_2 > \lambda_2^0$, again contradicting the assumption.
\end{description}
This shows that a fused group $F_i \cup F_j$ has to remain fused for an interval $[\lambda_{2}^{0},\lambda_{2}^{1}]$. However, for our algorithm it is necessary that then a flow exists for which all flows coming from the source are at maximum capacity. 
As the grouping with $F_{i}\cup F_{j}$ is optimal for $[\lambda_{2}^{0},\lambda_{2}^{1}]$,
using the subgradient equations  (\ref{eq:subgradientGeneralFusedLasso})
as well as the linearity of $\bbeta^{\mathcal{A}}$ and $\bs$ in $\lambda_{2}$
and that $t_{kl}=\text{const.}$ for $(k,l)\in E$, $k\in F_{i}\cup F_{j}$
and $l\not\in F_{i}\cup F_{j}$ we have that the $\tau_{kl}$ for
$(k,l)\in E;k,l\in F_{i}\cup F_{j}$ have to satisfy a linear equation
of the form 
\[
\sum_{(k,l)\in E;l\in F_{i}\cup F_{j}}\tau_{kl}+\lambda_{2}c_{k}=d_{k} \quad \text{for} \quad k\in F_{i}\cup F_{j}
\]
for some vectors $c_{k}$ and $d_{k}$. As the grouping is optimal,
we also know that a solution exists for every $\lambda_{2}\in[\lambda_{2}^{0},\lambda_{2}^{1}]$.
The space of all solutions of this linear equation is a vector
space and thus we know that there exist linear functions $\tau_{kl}(\lambda_{2})$
that satisfy these equations. But then, the flows $f_{kl}=\partial\tau_{kl}/\partial\lambda_{2}$
satisfy the maximum flow problem of the graph $\tilde{\mathcal{G}}_{F_{i}\cup F_{j}}$
with all flows from the source at maximum capacity (as the maximum
flow problem solves the above equations after taking the derivative
w.r.t. $\lambda_{2}$). Therefore, the group $F_{i}\cup F_{j}$ will
not be forced to break up by our algorithm.

Now the only thing that remains to do is to make sure that the rules
for activating and deactivating sets are correct. In order to see
this, we again use the continuity and piecewise linearity of $\bbeta$
w.r.t. $\lambda_{2}$. It is easy to see, that this also implies continuity
and piecewise linearity for $\bs$ w.r.t. $\lambda_{2}$.

\begin{description}
\item [{Activate~an~inactive~set:}] Let $F_{i}$ be the inactive where
w.l.o.g. $s_{F_{i}}(\lambda_{2}^{0})=1$ as well as $\beta_{F_{i}}(\lambda_{2}^{0})=0$
and $(\partial s_{F_{i}}/\partial\lambda_{2})(\lambda_{2}^{0})>0$.
First, the set cannot remain inactive for $\lambda_{2}>\lambda_{2}^{0}$,
as then we would have $(\partial s_{F_{i}}/\partial\lambda_{2})(\lambda_{2}^{0})\le0$,
which is a contradiction. Also $(\partial\beta_{F_{i}}/\partial\lambda_{2})(\lambda_{2})<0$
is not possible, as then $s(\lambda_{2})=-1$ for $\lambda_{2}>\lambda_{2}^{0}$,
which would violate the continuity of $\bs$. Therefore, 
$(\partial\beta_{F_{i}}/\partial\lambda_{2})(\lambda_{2})>0$
and thus $\beta_{F_{i}}(\lambda_{2})>0$ for $\lambda_{2}>\lambda_{2}^{0}$
and then $F_{i}$ satisfies the activity condition.
\item [{Deactivate~an~active~set:}] Let $F_{i}$ be active where w.l.o.g.
$s_{F_{i}}(\lambda_{2}^{0})=1$, $\beta_{F_{i}}(\lambda_{2}^{0})=0$
and $(\partial\beta_{F_{i}}/\partial\lambda_{2})(\lambda_{2}^{0})<0$.
Then due to the negative derivative, we have $\beta_{F_{i}}(\lambda_{2})\le0$
for $\lambda_{2}>\lambda_{2}^{0}$. However, if $\beta_{F_{i}}(\lambda_{2})<0$,
then $s_{F_{i}}(\lambda_{2})=-1$, again violating the continuity
of $\bs$. Thus, we have that $\beta_{F_{i}}(\lambda_{2})=0$ for $\lambda_{2}\in(\lambda_{2}^{0},\lambda_{2}^{0}+\varepsilon)$
and therefore, the condition for being active is not satisfied for
$F_{i}$. Thus it is inactive.
\end{description}
\end{proof}

\bibliographystyle{plainnat}
\bibliography{hhoeflin}